\documentclass[acmsmall]{acmart}

\usepackage{listings}
\usepackage{xcolor}
\usepackage{multirow}
\usepackage{graphics}
\usepackage{tabularray}
\usepackage{tabularx}
\usepackage{pifont}
\usepackage{amsmath}
\usepackage{tcolorbox}
\usepackage{mdframed,lipsum}
\usepackage{subcaption}
\usepackage{wrapfig}

\usepackage{xparse}
\usepackage{xspace}
\usepackage{comment}
\usepackage{soul}
\usepackage{colortbl}
\usepackage{pdfpages}
\usepackage[T1]{fontenc}

\DeclareTextFontCommand{\mytexttt}{\ttfamily\hyphenchar\font=45\relax}

\newcolumntype{P}[1]{>{\centering\arraybackslash}p{#1}}

\definecolor{red_bug}{HTML}{edcbc6}
\definecolor{green_a}{HTML}{a2bca3}
\definecolor{green_b}{HTML}{c6dbc1}
\definecolor{green_c}{HTML}{e9f1e6}
\definecolor{green_d}{HTML}{eff1d7}

\definecolor{codegreen}{rgb}{0.13,0.41,0.24}
\definecolor{codegray}{rgb}{0.5,0.5,0.5}
\definecolor{codepurple}{rgb}{0.52,0.03,0.40}
\definecolor{backcolour}{rgb}{1.0,1.0,1.0}

\definecolor{candyapplered}{rgb}{1.0, 0.03, 0.0}
\definecolor{lightmagenta}{HTML}{E1BEE7}
\definecolor{lightpurple}{HTML}{C5CAE9}
\definecolor{lightblue}{HTML}{B3E5FC}
\definecolor{lightcyan}{HTML}{B2DFDB}

\definecolor{textgreen}{HTML}{2CB32C}
\definecolor{textred}{HTML}{DC143C}

\usepackage{enumitem, kantlipsum}
\usepackage{filecontents,lipsum}
\usepackage{graphicx}
\usepackage{multirow}
\usepackage[linesnumbered,ruled,vlined]{algorithm2e}

\SetCommentSty{mycommfont}

\SetKwInput{KwInput}{Input}                
\SetKwInput{KwOutput}{Output}              

\newcommand{\iFix}{\textsc{iFix}\xspace}

\newcommand{\circled}[1]{{\large \textcircled{\small #1}}}

\newcommand{\CodeIn}[1]{{\small\mytexttt{#1}}}

\newcommand{\CodeInTable}[1]{\footnotesize\mytexttt{#1}}
\newcommand{\Comment}[1]{}

\newcommand{\NumUserStudyParticipants}{28\xspace}
\newcommand{\SRankingBoost}{{45}\%\xspace}
\newcommand{\LRankingBoost}{{93}\%\xspace}
\newcommand{\MCRGenBoost}{{39}\%\xspace}
\newcommand{\CUREBoost}{{51}\%\xspace}
\newcommand{\UserSuccessRateBoostAgainstManual}{{40}\%\xspace}
\newcommand{\UserSuccessRateBoostAgainstBaseline}{{8}\%\xspace}
\newcommand{\UserDebugTimeReducedAgainstManual}{{36}\%\xspace}
\newcommand{\UserDebugTimeReducedAgainstBaseline}{{33}\%\xspace}
\newcommand{\UserPreferenceBoostAgainstManual}{{50}\%\xspace}
\newcommand{\UserPreferenceBoostAgainstBaseline}{{33}\%\xspace}
\newcommand{\UserConfidenceBoostAgainstManual}{{50}\%\xspace}
\newcommand{\UserConfidenceBoostAgainstBaseline}{{20}\%\xspace}

\newcommand{\iFixFrontEndLOC}{1,050\xspace}
\newcommand{\iFixBackEndLOC}{1,690\xspace}
\newcommand{\iFixInstruLOC}{3200\xspace}

\newcommand{\credit}[1]{#1}

\newcommand{\distance}{8pt}
\newcounter{finding}
\newenvironment{finding}[1]
{
    \refstepcounter{finding}
	\begin{mdframed}[
    	nobreak=true,
    	linecolor=black,
    	roundcorner=12pt,
    	backgroundcolor=gray!01,
    	linewidth=0.5pt,
    	leftmargin=0.5em,
    	rightmargin=0.5em,
    	topline=true,
    	bottomline=true,
    	frametitlerule=true,
    	frametitlebackgroundcolor=gray!20,
    	frametitlerulecolor=gray,
    	frametitle=Answer to RQ\arabic{finding},
    	frametitleaboveskip=0.3em,
    	frametitlebelowskip=0.2em,
    	skipabove=12pt
	]
}
{
    \end{mdframed}
    \vspace{1em}
}

\setcopyright{cc}
\setcctype{by}
\acmDOI{10.1145/3720510}
\acmYear{2025}
\acmJournal{PACMPL}
\acmVolume{9}
\acmNumber{OOPSLA1}
\acmArticle{145}
\acmMonth{4}
\received{2024-10-16}
\received[accepted]{2025-02-18}

\begin{document}

\title[Saving 1/3 of Debugging Time in Program Repair with Interactive Runtime Comparison]{Show Me Why It’s Correct: Saving 1/3 of Debugging Time in Program Repair with Interactive Runtime Comparison}

\author{Ruixin Wang}
\orcid{0009-0002-9298-0925}
\affiliation{%
  \institution{Purdue University}
  \city{West Lafayette}
  \country{USA}
}
\email{ruixinw@purdue.edu}

\author{Zhongkai Zhao}
\orcid{0000-0003-2365-9898}
\affiliation{%
  \institution{National University of Singapore}
  \city{Singapore}
  \country{Singapore}
}
\email{zhongkai.zhao@u.nus.edu}

\author{Le Fang}
\orcid{0009-0005-1351-5681}
\affiliation{%
  \institution{Purdue University}
  \city{West Lafayette}
  \country{USA}
}
\email{fang250@purdue.edu}

\author{Nan Jiang}
\orcid{0000-0001-8518-2576}
\affiliation{%
  \institution{Purdue University}
  \city{West Lafayette}
  \country{USA}
}
\email{jiang719@purdue.edu}

\author{Yiling Lou}
\orcid{0000-0002-4066-3365}
\affiliation{%
  \institution{Fudan University}
  \city{Shanghai}
  \country{China}
}
\email{yilinglou@fudan.edu.cn}

\author{Lin Tan}
\orcid{0000-0002-6690-8332}
\affiliation{%
  \institution{Purdue University}
  \city{West Lafayette}
  \country{USA}
}
\email{lintan@purdue.edu}

\author{Tianyi Zhang}
\orcid{0000-0002-5468-9347}
\affiliation{%
  \institution{Purdue University}
  \city{West Lafayette}
  \country{USA}
}
\email{tianyi@purdue.edu}



\begin{abstract}

Automated Program Repair (APR) holds the promise of alleviating the burden of debugging and fixing software bugs. Despite this, developers still need to manually inspect each patch to confirm its correctness, which is tedious and time-consuming. This challenge is exacerbated in the presence of plausible patches, which accidentally pass test cases but may not correctly fix the bug. To address this challenge, we propose an interactive approach called \iFix to facilitate patch understanding and comparison based on their runtime difference. \iFix performs static analysis to identify runtime variables related to the buggy statement and captures their runtime values during execution for each patch. These values are then aligned across different patch candidates, allowing users to compare and contrast their runtime behavior. To evaluate \iFix, we conducted a within-subjects user study with \NumUserStudyParticipants participants. Compared with manual inspection and a state-of-the-art interactive patch filtering technique, \iFix reduced participants' task completion time by \UserDebugTimeReducedAgainstManual and \UserDebugTimeReducedAgainstBaseline while also improving their confidence by \UserConfidenceBoostAgainstManual and \UserConfidenceBoostAgainstBaseline, respectively. Besides, quantitative experiments demonstrate that \iFix  improves the ranking of correct patches by at least \MCRGenBoost compared with other patch ranking methods and is generalizable to different APR tools. 

\end{abstract}






\maketitle
\section{Introduction}
\label{sec:intro}

\noindent Bug fixing is a common but challenging task in software development. A recent survey with 950 developers shows that 44\% of developers identified bug fixing as their biggest pain point in software development~\cite{2021survey}. Automated Program Repair (APR) holds the promise to address this challenge by automating or augmenting the manual bug-fixing process. Over the past decade, various APR techniques have been proposed, spanning across template-based~\cite{tbar,simfix}, search-based~\cite{sun2018search}, semantic-based~\cite{le2017s3}, heuristic-based~\cite{arja-heuristic-1,heuristic-2,elixir-heuristic-3}, constraint-based~\cite{ACS-constraint-1, nopol-constraint-2, constraint-3}, and learning-based~\cite{chen_sequencer_2019, lutellier_coconut_2020, Li2020dlfix, jiang_cure_2021, xia2024automated} methods.

Although developers are open to utilizing patches created by APR techniques, they are skeptical of the quality of auto-generated patches~\cite{winter_how_2022, zhang_program_2022, Eladawy24icse}. As a result, they still need to manually \credit{\textit{comprehend},  \textit{validate}, and \textit{compare}} candidate patches to answer the question: \textit{\textbf{why is this patch correct?}} Liang et al.~found that, on average, developers spent 23 minutes manually \credit{inspecting and comparing} candidate patches generated by APR~\cite{liang_interactive_2021}. This issue is exacerbated by the presence of {\em plausible patches} that pass the test cases but may not fully fix a bug or address the root cause of a bug~\cite{long_analysis_2016, smith_is_2015, le_overfitting_2018}. Eladawy et al.~highlighted this concern in their user study: Compared to manually debugging without candidate patches, receiving a plausible but incorrect patch decreases participants’ success rate by 65\%~\cite{Eladawy24icse}.

There are two \textit{key challenges} to help developers validate patch correctness. \textbf{First, APR tools often generate many plausible patches}. According to Noller et al.~\cite{noller_trust_2022}, 72\% of developers did not want to review more than five patches and quickly lost confidence in APR if the first few patch candidates were incorrect. \textbf{Second, it is tedious and time-consuming to \credit{comprehend and} validate whether a patch fully fixes the bug or accidentally passes the test cases.} Given the large amount of runtime information and the complexity of call stack traces in many real programs, it is cumbersome to step through program statements with a patch applied, not to mention comparing multiple patch candidates. In practice, a developer needs to compare the runtime behavior of multiple execution traces and switch back and forth to understand their behavior difference. \credit{Cambronero et al.~observed in their user study that ``{\em developers often spent time investigating the roles that variables from the generated patches play in the original defective code}'' and ``{\em they spent a lot of time trying to find contextual information to assess the correctness of the given candidate patches}'' \cite{vlhcc}}. 

To address the first challenge, many patch ranking and classification techniques \cite{directfix, le2017s3, long2016automatic, saha_elixir_2017, shibboleth, xia_less_2022, ye_2021_automated, xiong2018identifying, varfix} have been proposed to prioritize patches that are more likely to be correct. However, developers still need to manually check and compare candidate patches in the ranked list, at least the top candidates, one by one to verify their correctness. Some recent techniques prompt developers with questions and filter candidate patches based on their responses \cite{gao2020interactive, bohme_2020_human, liang_interactive_2021}. However, they only rely on a pre-defined set of questions, which are limited in scope and unable to comprehensively capture runtime behavior differences, as candidate patches often vary at specific points that such questions may fail to address. An alternative way is to cluster patch candidates, so developers only need to inspect one patch per cluster ~\cite{cashin2019understanding, martinez2024test}. Yet there can be many clusters depending on the specificity of patch similarity metrics and thresholds. These patch clustering methods do not provide any support for navigating through the clusters and comparing patches from different clusters. 

Notably, none of the existing approaches help developers fully validate a plausible patch, e.g., examining how the erroneous runtime values originate from the buggy line and propagate to the test oracle, how the plausible patch alters the runtime behavior to fix the bug, etc. This is a critical step in the real-world adoption of APR techniques. As shown by Cambronero et al.~\cite{vlhcc}, developers still spent significant time stepping through the patched program line by line \credit{to understand how its execution differs from the execution of the buggy program}. As a result, their productivity did not improve much compared to those participants who were asked to fix the given bug manually without the APR-generated patches. 

To bridge this gap, we propose a new interactive approach called \iFix to help developers better understand the runtime behavior of auto-generated patches. To address the first challenge, \iFix uses hierarchical clustering to group similar patches and provides interactive support for developers to navigate patches with different levels of similarity. Developers can easily filter the clusters of undesired patches and zoom into clusters with promising ones.  
To address the second challenge, \iFix employs static analysis to identify the runtime values \credit{of \textit{variables}, \textit{method calls}, and \textit{subexpressions}} relevant to the patched statements, align them across different patch candidates, and highlight the runtime differences via color-coding. This helps users understand how the erroneous values are propagated through the program and how these values are fixed by different patches. We have implemented \iFix as a VS Code extension. 

To evaluate the usefulness and usability of \iFix, we first conducted a within-subjects user study with \NumUserStudyParticipants developers. We compared \iFix with (1)  \credit{manually inspecting candidate patches} and (2) InPaFer~\cite{liang_interactive_2021}, a state-of-the-art interactive patch filtering technique. Compared with these two baselines, \iFix improves the bug fixing success rate by  \UserSuccessRateBoostAgainstManual and \UserSuccessRateBoostAgainstBaseline while reducing the bug fixing time by \UserDebugTimeReducedAgainstManual and \UserDebugTimeReducedAgainstBaseline, respectively. \iFix also notably enhances the users' confidence in the final patch by \UserConfidenceBoostAgainstManual and \UserConfidenceBoostAgainstBaseline respectively. Finally, our additional experiments show that \iFix helps developers inspect fewer candidate patches by improving the overall ranking of correct patches by ~\SRankingBoost and ~\LRankingBoost compared to a static-feature-based patch ranking technique~\cite{le2017s3} and a learning-based ranking method~\cite{xia_less_2022}, respectively. Further experiments on other APR techniques~\cite{rewardrepair, knod} demonstrate that our approach is generalizable to different APR techniques.

Overall, our work makes the following contributions:

\begin{itemize}
  \item We introduced \iFix, an interactive approach to help developers better understand the runtime behavior of auto-generated patches and boost their performance of bug fixing.
  
  \item We implemented \iFix as a VS Code plugin and have open-sourced our implementation to foster future research \cite{ifix_github, ifix_zenodo}.
  
  \item We conducted a user study and a series of benchmark experiments to evaluate the usability, effectiveness, and generalizability of \iFix. We have released our data for reproducibility.\footnote{\label{note2}\href{https://sites.google.com/view/ifixuserstudy}{https://sites.google.com/view/ifixuserstudy}}
\end{itemize}


The rest of the paper is organized as follows. Section \ref{sec:MotivatingExample} motivates \iFix with a real-world example. Section \ref{sec:Approach} describes the approach. Sections \ref{sec:UserStudy} and \ref{sec:QuantitativeExperiment} describe the user study and benchmark experiments to evaluate \iFix. Section \ref{sec:Discussion} discusses design implications, threats to validity, and future work. Section \ref{sec:RelatedWork} discusses related work. Section \ref{sec:Conclusion} concludes the paper.

\section{Motivating Example}
\label{sec:MotivatingExample}
\lstdefinestyle{code}{
  backgroundcolor=\color{backcolour},
  commentstyle=\color{codegreen},
  keywordstyle=\color{codepurple},
  numberstyle=\tiny\color{codegray},
  stringstyle=\color{magenta},
  basicstyle=\ttfamily\tiny,
  breakatwhitespace=false,         
  breaklines=true,                 
  captionpos=b,                    
  keepspaces=true,                 
  numbers=left, 
  showspaces=false,                
  showstringspaces=false,
  showtabs=false,                  
  tabsize=2,
  escapeinside=||
}

\lstset{numbers=left, numberblanklines=false, escapeinside=||, style=code}
\let\origthelstnumber\thelstnumber
\makeatletter
\newcommand*\Suppressnumber{%
  \lst@AddToHook{OnNewLine}{%
    \let\thelstnumber\relax%
     \advance\c@lstnumber-\@ne\relax%
    }%
}

\newcommand*\Reactivatenumber[1]{%
  \setcounter{lstnumber}{\numexpr#1-1\relax}
  \lst@AddToHook{OnNewLine}{%
   \let\thelstnumber\origthelstnumber%
   \refstepcounter{lstnumber}
  }%
}

\makeatother

\makeatletter
\newcommand{\customlabel}[2]{%
   \protected@write \@auxout {}{\string \newlabel {#1}{{#2}{\thepage}{#2}{#1}{}} }%
   \hypertarget{#1}{}
}
\makeatother

\begin{figure}[t]
    \centering
    \customlabel{code-a}{1a}
    \customlabel{code-b}{1b}
    \customlabel{code-c}{1c}
    \includegraphics[width=\linewidth]{Figures/codesnippet-w.pdf}
    \caption{Code snippets of a bug in Apache Commons Math~\cite{bug_page}}
    \label{fig:codebase-1}
\end{figure}

\noindent This section illustrates the usage of \iFix with a real-world bug from the Apache Commons Math project~\cite{bug_page}. The bug arises in the method \CodeIn{MannWhitneyUTest.calculateAsymptoticPValue}, which computes the asymptotic p-value using the normal approximation. Figure~\ref{fig:codebase-1} shows the call stack from the failed test case (Figure~\ref{code-c}) to the buggy method (Figure~\ref{code-a}). This test case is designed to test \CodeIn{MannWhitneyUTest} with two data samples, each containing 1500 integers. The test fails because the computed p-value exceeds the expected threshold of $0.1$ (Line \#115, Figure~\ref{code-c}). The buggy statement is located at Line \#173 in method \CodeIn{calculateAsymptoticPValue} (Figure ~\ref{code-a}), which calculates the product of two integers.


Suppose Alice is an open-source developer for Apache Commons Math, and she is assigned to fix this bug. 
The code in \CodeIn{MannWhitneyUTest.java} was written by a developer who is no longer active in the open-source community. Since Alice is unfamiliar with the underlying statistics of the Mann-Whitney U Test, she decides to try an APR tool called CURE~\cite{jiang_cure_2021}\footnote{We use CURE~\cite{jiang_cure_2021} as an example, but it can be any APR tool. Our approach is designed to be agnostic to any APR tools.} and check whether it can generate any inspiring patches to help her fix this bug \credit{before attempting to fix it manually}. CURE generates thousands of patches, including 33 plausible patches that pass all test cases, as shown in Table~\ref{table: example-patches}. All these 33 patches fix the test failure in \CodeIn{MannWhitneyUTestTest.testBigDatasSet} without introducing another failure. Alice is impressed, but she also realizes that some patches seem to fix the bug simply by coincidence after a glance.

\begin{table}[h]
\centering
\footnotesize
  {%
    \caption{The original ranking of the plausible patches generated by CURE~\cite{jiang_cure_2021} ($\star$ indicate the correct patch)}
    \vspace{-6pt}
    \label{table: example-patches}
  \begin{tabular}{rc}
    \toprule
    Index & APR-Generated Patches\\
    \midrule
    1\ \ \  & \CodeInTable{double n1n2prod = n1*( n2+ n2) /2.0;}\\
    2\ \ \  & \CodeInTable{double n1n2prod = n1*n2;}\\
    3\ \ \  & \CodeInTable{final int n1n2prod =2*( n1+ n2+1);}\\
    4\ \ \  & \CodeInTable{double n1n2prod = n2*( n1+ n2+1) /2.0;}\\
    5\ \ \  & \CodeInTable{final double n1n2prod = n1*( n1+2+1) /2.0;}\\
    $\star$ 6\ \ \  & \CodeInTable{final double n1n2prod = n1*n2;}\\
      & ...  \\
    33\ \  & \CodeInTable{int n1n2prod = n1*2*2;} \\
    \bottomrule
  \end{tabular}%
  }
\end{table}

\textbf{{The Pain of Manual Inspecting Plausible Patches:}} \credit{To identify the correct patch among the candidates, Alice needs to understand how each patch eliminates the bug through debugging. However, this process is tedious and time-consuming, especially since many patches look very similar to each other and only have subtle differences. In some cases, it may be more efficient to fix the bug herself rather than rely on auto-generated patches.}

\credit{Still, exploring auto-generated patches can be beneficial, especially when there is significant uncertainty about the root cause of the bug. If Alice decides to evaluate the candidate patches, determining the correct fix remains a challenge.} Though she may not have to investigate all 33 plausible patches, debugging several promising candidates already takes a long time. 
\credit{For each patch, Alice needs to analyze its impact on the program’s execution. While some differences between patches may be noticed by simply scanning the code, fully understanding how a patch affects runtime behavior often requires setting breakpoints and stepping through program statements.} Alice also needs to compare the runtime behavior of multiple patch candidates and decide which one truly fixes the bug.

\begin{figure*}
  \customlabel{teaser-1}{2\circled{1}}
  \customlabel{teaser-2}{2\circled{2}}
  \customlabel{teaser-1a}{2.1a}
  \customlabel{teaser-2a}{2.2a}
  \customlabel{teaser-2b}{2.2b}
  \customlabel{teaser-2c}{2.2c}
  \customlabel{teaser-I}{2.I}
  \customlabel{teaser-II}{2.II}
  \customlabel{teaser-III}{2.III}
  \includegraphics[width=\textwidth]{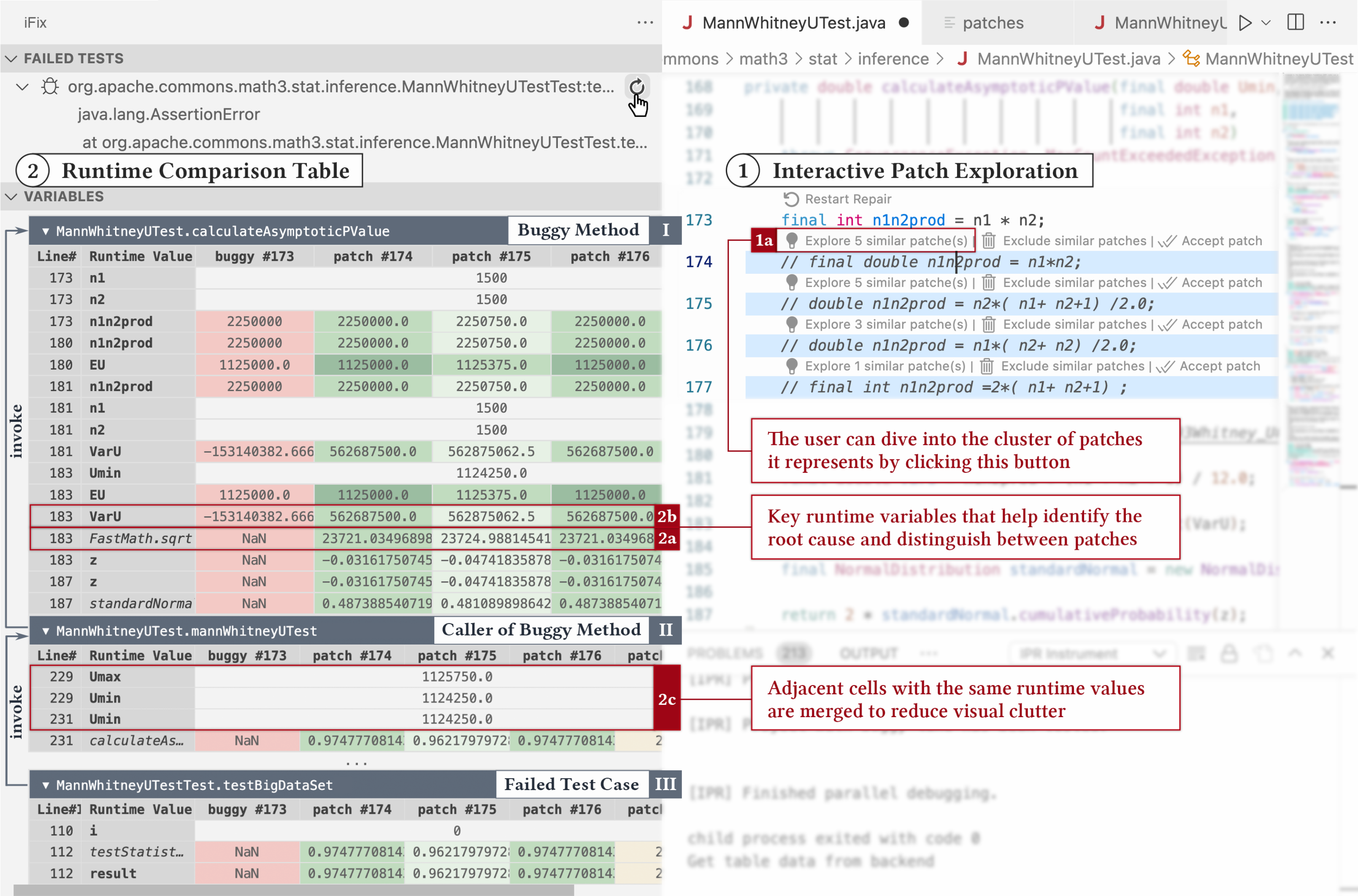}
  \vspace{-1em}
  \caption{The User Interface of \iFix.}
  \label{fig:teaser}
\end{figure*}

\textbf{{How \iFix Can Help:}} Alice decides to use {\iFix} to explore the 33 patches. {Given that manually checking 33 patches is overwhelming,} \iFix first selects a small yet diverse set of patches, as shown in Figure~\ref{teaser-1}. 
\credit{None of the existing tools can efficiently compare the runtime behavior of multiple patches at once, making it difficult to understand how each patch attempts to fix the bug without manually setting breakpoints and stepping through statements. 
\iFix eliminates this hassle by} providing a series of {\em Runtime Comparison Tables} with runtime values related to the buggy statement and patches (Figure~\ref{teaser-2}). Each of these tables corresponds to a method in the call stack shown in Figure~\ref{fig:codebase-1}. For example, the table in Figure~\ref{teaser-I} corresponds to the buggy method \CodeIn{calculateAsymptoticPValue} (Figure~\ref{code-a}). Figure \ref{teaser-II} corresponds to the method {mannWhitneyUTest} (Figure~\ref{code-b}) that invokes \CodeIn{calculateAsymptoticPValue}, and Figure \ref{teaser-III} corresponds to the method that invokes the second one, which is the failed test case \CodeIn{testBigDataSet} (Figure~\ref{code-c}). 

In each table, \iFix documents the method's execution trace by logging the runtime values of the variables used or affected by the patch, \credit{as well as the values of method calls and subexpressions involving the identified variables}. The first two columns, \textsf{Runtime Value} and \textsf{Line\#}, show the names of \credit{variables, method calls, and subexpressions} along with their corresponding line numbers in the source code. The following columns show their values under the buggy code and the code applied with each candidate patch, respectively. With such a holistic view of runtime behavior across all versions of the program, Alice can now compare them side-by-side without having to debug each patch separately.

While examining the runtime values, Alice notices a significant difference in the return values of \CodeIn{FastMath.sqrt(VarU)} between the buggy code and all plausible patches (Figure~\ref{teaser-2a}). This immediately stands out because \CodeIn{NaN} produced by the buggy code is an \textit{unusual and invalid} result for a mathematical operation. In contrast, all the patches produce real numbers. Additionally, Alice observes that the variable \CodeIn{VarU} is negative in the buggy code but positive in all the patches (Figure~\ref{teaser-2b}). Since negative numbers cannot have a square root, Alice realizes that the \CodeIn{NaN} in the buggy code is caused by the negative value of \CodeIn{VarU}.

Alice then inspects the code to understand why \CodeIn{VarU} ended up with a negative value in the buggy code. She finds that on Line 181 of the buggy method (Figure~\ref{code-a}), \CodeIn{VarU} is computed from the expression \CodeIn{n1n2prod * (n1 + n2 + 1) / 12.0}. According to the runtime comparison table (Figure~\ref{teaser-I}), \CodeIn{n1n2prod}, \CodeIn{n1}, and \CodeIn{n2} are all positive numbers, so \CodeIn{VarU} should theoretically be positive as well. However, Alice notices that the runtime values of \CodeIn{n1n2prod} are extremely large (e.g. $2250,000$). She realizes that \textit{an integer overflow occurs when computing \CodeIn{VarU}}, which leads to a negative value and ultimately causes the test case to fail. She needs to use a different data type rather than \CodeIn{int}, as \CodeIn{int} in Java is limited to 4 bytes, with a maximum value of $2147483647$.

Alice then takes a closer look at the candidate patches (Figure~\ref{teaser-1}). She notices that Patch \#174 stands out because it only changes the type of the variable \CodeIn{n1n2prod}, while the other patches alter the expression on the right-hand side. Given her understanding of the bug's root cause, which is an integer overflow, Alice concludes that modifying the data type is sufficient, and there’s no need to adjust the expression itself. Therefore, she identifies Patch \#174 as the most promising option. With the help of \iFix, Alice quickly selects this patch as the correct fix from the set of patches generated by the APR tools. She is confident in her choice, as \iFix has guided her in pinpointing the root cause of the bug.

\section{Approach}
\label{sec:Approach}
In this section, we first describe the design principles and give an overview of the user interface and interactive features of \iFix. Then, we explore the algorithms and technical details for each component in the following subsections.

\subsection{Design Principles}

The design of \iFix is guided by empirical findings on the difficulties developers face with Automated Program Repair (APR) \cite{noller_trust_2022, winter_lets_2023, winter_how_2022, zhang_program_2022, vlhcc}. To address the two key challenges (mentioned in Section \ref{sec:intro}) on the validation of patch correctness, a tool should assist developers in the following aspects:

\begin{enumerate}
    \setlength\itemsep{0.5mm}
    \item \textit{Reduce the Number of Patches to Investigate.} Studies show that reviewing many patches leads to cognitive overload, making patch validation inefficient \cite{winter_lets_2023, noller_trust_2022}. Most developers prefer reviewing no more than \textit{five patches}~\cite{noller_trust_2022}.
    
    \item \textit{Provide Concrete Evidence for Patch Correctness.} Developers do not fully trust APR-generated patches when validation relies solely on passing test cases \cite{winter_how_2022, zhang_program_2022}. Instead, they expect additional evidence, such as execution traces and runtime behavior \cite{noller_trust_2022, vlhcc}.
    
    
    \item \textit{Enable Holistic Patch Comparison.} Studies reveal that developers often struggle to compare multiple patches, frequently relying on ad-hoc approaches \cite{vlhcc}. Such a challenge calls for holistic patch comparison to improve the efficiency of patch validation and selection
\end{enumerate}

\subsection{Overview}
\label{approach_overview}
\noindent Figure \ref{fig:teaser} shows the user interface of \iFix, which has been integrated into VS Code as an extension. \iFix includes two main components---\textit{interactive patch exploration} and \textit{runtime behavior comparison}. 

\textbf{Interactive Patch Exploration.} A recent large-scale survey~\cite{noller_trust_2022} shows that 72\% of developers did not want to review more than five patches and can quickly lose confidence in APR if the first few patch candidates are not correct. To mitigate this challenge, \iFix allows users to navigate patches with different levels of similarity through patch exploration (Figure \ref{teaser-1}). It renders a short list of representative and diverse patches selected by the {\em patch clustering and sampling} algorithm (Section~\ref{subsec:multi-criteria}).  For each patch, a user can dive into the cluster of patches it represents by clicking the {``Explore Similar Patches''} button over it (Figure~\ref{teaser-1a}). This feature is useful in cases where the initially selected patch looks promising but is not fully correct, since users can explore similar patches in the same cluster and locate the correct one. On the other hand, if a patch looks completely wrong, users can reject it and similar ones which are likely to be wrong, by clicking the {``Exclude Similar Patches''} button. Using these interactive features together, the user can easily navigate a large number of patches and locate the patch that looks the most promising to her. 

\vspace{2pt} 

\textbf{Runtime Comparison Table.} The same survey~\cite{noller_trust_2022} also shows that 69\% of developers wished to see additional information such as runtime values to validate auto-generated patches. \credit{Another survey~\cite{vlhcc} reveals that the absence of a holistic patch comparison mechanism prevents developers from investigating patch candidates efficiently.}
To help users understand how the erroneous values are propagated among the statements and how these values are fixed by different patches, we propose a {\em runtime behavior comparison} method,  which performs static analysis to identify runtime variables related to the patched statements, captures the runtime values, and aligns them in a series of runtime comparison tables (Section~\ref{subsec:comparison}). Figure \ref{teaser-2} shows the runtime comparison tables. Each comparison table corresponds to the execution trace of a method call in the call stack. Specifically, the first table (Figure~\ref{teaser-I}) corresponds to the method in which the buggy statement occurs, and the last table (Figure~\ref{teaser-III}) corresponds to the failed test case. 

These tables simulate the debugging processes applied with several candidate patches simultaneously, thereby aiming to minimize programmers' efforts of debugging during patch selection. A user can collapse the tables for those less important method calls to simplify the user interface. Also, when the user clicks on a variable name in the table, \iFix will automatically redirect the user to the line of code that computes the specific value of that variable in the code editor.

In each runtime comparison table, Column \textsf{Line\#} shows the line number where the runtime value is logged. Column \textsf{Runtime Value} shows the name of the runtime value, which can be a variable name, a method call, or a subexpression. The following columns display the runtime value logged from the original buggy code and from individual patches. 

To help developers easily see the differences in runtime values, \iFix color-codes each unique value in a row. As shown in Figure \ref{teaser-2}, if the buggy code has a different runtime value than the patches, the corresponding cell is rendered in \colorbox{red_bug}{\footnotesize\texttt{red}}, while cells with unique values in the patches are rendered in different shades of green (\xspace\colorbox{green_a}{\rule[1pt]{0pt}{0.3em}\xspace\xspace\xspace}\colorbox{green_b}{\rule[1pt]{0pt}{0.3em}\xspace\xspace\xspace}\colorbox{green_c}{\rule[1pt]{0pt}{0.3em}\xspace\xspace\xspace}\colorbox{green_d}{\rule[1pt]{0pt}{0.3em}\xspace\xspace\xspace}\ ). Furthermore, to reduce the visual clutter of rendering many runtime values, \iFix combines adjacent cells with the same runtime value. For example, in Figure~\ref{teaser-2c}, the cells with identical values for the variables \CodeIn{Umax} and \CodeIn{Umin} are combined.

Finally, to distinguish runtime values logged from different loop iterations or repeated method calls, \iFix adds the index of the loop iteration or method call after the line number, separated by a \#. For subexpressions and method calls that vary across patches, \iFix synthesizes a new name to align their values in a single row. Specifically, \iFix retains the common substrings and substitutes the variations with a wildcard symbol *. For example, in Figure~\ref{fig:ui-preview-2-sub1}, the patches substitute the second argument of \CodeIn{Character.codePointAt} with different values. \iFix synthesizes a new name, \CodeIn{Character.codePointAt} \CodeIn{(input, *)} to represent this method call, so it can render the return value of this method call from different patches in the same row, as shown in Figure~\ref{fig:ui-preview-2-sub2}.

In the following subsections, we describe the algorithms and implementation details behind these two key features. 

\vspace{-0.5em}

\subsection{Patch Clustering and Sampling}
\label{subsec:multi-criteria}
\noindent \iFix applies hierarchical clustering and multi-criteria sampling to select a small set of patches that are diverse and representative. This helps developers interactively examine a small number of patches each time, provide feedback, and gradually narrow down to the correct one in a large patch space. Compared with other clustering algorithms such as K-Means~\cite{K-means} and KNN~\cite{KNN}, hierarchical clustering is not sensitive to initialization conditions such as the random seeds and the original ranking. Furthermore, compared with K-Means and KNN, there is no need to specify the number of clusters beforehand.

\begin{algorithm}[h]
\caption{Patch Clustering and Sampling}
\label{algorithm: multi-criteria}
\small
\DontPrintSemicolon
  
  \KwInput{Plausible Patch List $ P $, Target Buggy Code $ B $}
  \KwOutput{Sampled Patch List $ L $}
  
  $ UL \leftarrow \{\}  $
  
    $ C \leftarrow Hierarchical\_Cluster(P) $

    \For{$ c $ in $ C $}
    {
      $ centroid \leftarrow \arg\max_{p \in c} R(p)$
    
      $ UL.append(centroid)  $
    }
  
  $ L \leftarrow Rank\_By\_Similarity(UL, B) $

\end{algorithm}

Algorithm \ref{algorithm: multi-criteria} describes the clustering and sampling process. First, \iFix performs hierarchical clustering~\cite{Hierarchical} to cluster the given patches based on their similarity (Line 2). \credit{Hierarchical clustering produces a dendrogram of the given patches. To avoid overwhelming users, {\iFix} selects the hierarchical level with no more than five clusters and cuts the dendrogram into clusters.} 
For each produced cluster, \iFix selects the most representative patch by selecting the \textit{centroid} of the cluster (Line 4). We use Levenshtein distance to measure the distance between a patch to another patch, since it effectively captures the syntactic differences between patches by assessing how closely one patch aligns with another ~\cite{UseLevenshteinDistance}. 
Finally, \iFix ranks the representative patches selected in the previous step to prioritize the correct patch (Line 6 in Algorithm \ref{algorithm: multi-criteria}). Our ranking heuristic is based on an insight that has been observed in several prior work~\cite{d2016qlose, le2017s3, wen2018context, wang2020automated}---{\em a correct patch is more likely to be similar to the buggy code}. 
For each selected patch, \iFix tokenizes it, calculates its Levenshtein distance to the tokenized buggy code, and ranks them based on the Levenshtein distance. We choose Levenshtein distance as the metric here for the same reasons as before—its effectiveness in capturing syntactic differences and subtle variations between patches.



\vspace{-0.5em}

\subsection{Runtime Behavior Comparison}
\label{subsec:comparison} 
\noindent Since the execution trace of a test failure contains many method invocations and runtime values, \iFix first performs {\em call graph analysis} and {\em def-use analysis} to identify a small set of runtime values related to the patched line of code. Then, \iFix performs {\em code instrumentation} to log the related runtime values. Finally, \iFix analyzes the logged values and aligns them across different patches in a set of tables for easy comparison. We elaborate on each step below.


\textbf{Call Graph Analysis.}
\iFix performs dynamic analysis to capture the call stack traces during execution. Specifically, \iFix leverages the \CodeIn{java.lang.getStackTrace} API to log the call stack trace of the patched statement. For example, the call stack trace in Figure~\ref{fig:codebase-1} includes the information about  \CodeIn{calculateAsymptoticPValues}, \CodeIn{mannWhiteneyUTest}, and \CodeIn{testBigDataSet}. The call stack trace helps narrow down the scope of def-use analysis in the next step. It enables programmers to understand how the buggy statement is triggered by the test case before patching. Additionally, it shows how faulty values propagate through the stack trace, leading to the failure.


\textbf{Def-Use Analysis.} Based on the call graph analysis result, \iFix performs def-use analysis on each method in the stack trace of the buggy statement. \credit{While def-use analysis is an established method, \iFix is the first to apply it to patch comprehension and validation.} Since there are many variables in each method, it would be overwhelming if \iFix renders all of them to programmers. Thus, this step aims to identify the variables related to the faulty value produced from the buggy statement in each method. 

A variable is considered {\em defined} if it is assigned a new value, e.g., appearing on the left-hand side of an assignment statement. A variable is considered {\em used} if its value is used to compute another value in an expression. Def-use analysis will return {\em def-use chains} for all variables defined in a method, as well as {\em use-def chains} for all variables used in a method. A def-use chain of a variable consists of the definition of the variable and all of its use locations reachable from that definition without any other intervening definitions. For example, the defined variable in the buggy statement in Figure~\ref{code-a} is \CodeIn{n1n2prod}. Its use locations include Line \#176 and Line \#177 in Figure \ref{code-a}. By contrast, a use-def of a variable consists of a location that uses the variable and all the variables that are directly computed from that variable in that location.

\begin{algorithm}[h]

\DontPrintSemicolon
\caption{Identify variables affected by a statement}
\label{algorithm: def-use}

\small
  \KwInput{Statement $s$, Method $m$}
  \KwOutput{Variables affected by the statement $s$}

  $DU\_Chains, UD\_Chains \gets Def\_Use\_Analysis{(m)}$ 

  $d \gets {Get\_Defined\_Variable}{(s, DU\_Chains)}$  

  $D.\text{append}(d)$

  $R \gets \{\}$

  \While{$D \neq \varnothing$}
    {$v \gets D.\text{pop}()$
    
    $R.\text{append}(v)$
    
    \For{$u \in DU\_Chains.\text{get}(v)$}
      {$D.\text{append}({UD\_Chains}.\text{get}(u))$
      
      $R \gets R.\text{append}(u)$
      
      }}
  $R \gets R \cup {DU\_Chains}.\text{get}(d)$
  
  \textbf{return} $R$
\end{algorithm}

Algorithm~\ref{algorithm: def-use} describes the procedure to gather runtime values affected by the buggy statement. After def-use analysis (Line 1), \iFix identifies the variable {\em defined} by the buggy statement (Line 2). Then, it analyzes the def-use chains of these variables and identifies all their used locations (Line 7). Given the use locations, it analyzes the use-def chains to identify the variables defined in those locations with the values computed from the buggy statement. For example, from the use locations (Figure~\ref{code-a}, \#180 - \#181) of the faulty value in \CodeIn{n1n2prod}, \iFix identifies that \CodeIn{EU} and \CodeIn{VarU} are further computed from this faulty value. \iFix repeats this step until no new variables are found (Lines 4-10). In this way, \iFix captures the propagation of a faulty value in a method. 

To continue tracing the propagation of the faulty values in the caller of the buggy method, \iFix runs the same procedure in Algorithm~\ref{algorithm: def-use} on the method invocation statement and continues to process other methods in the call stack. On the call site, \iFix identifies the arguments of the method call statements and keeps tracing them. For example, \iFix analyzes the variables by the call site at Line \#231 in \CodeIn{calculateAsymptoticPValue} (Figure \ref{code-b}) and keeps tracing variables \CodeIn{Umin}, \CodeIn{x.length}, and \CodeIn{y.length}. \CodeIn{x.length} and \CodeIn{y.length} are traced since they are passed in as arguments for \CodeIn{n1} and \CodeIn{n2}, both of which are used in the buggy statement. \CodeIn{Umin} is also traced since it is passed in as an argument for \CodeIn{Umin} which is used to compute \CodeIn{z}, given \CodeIn{z} is also involved in the propagation of the faulty values.
\iFix further analyzes the variables by the call site at Line \#113 (Figure \ref{code-c}) in \CodeIn{mannWhiteneyUTest} in a similar manner.


\textbf{Code Instrumentation.} Given the related variables identified from the previous step, \iFix instruments the defined and used locations of these variables to log their runtime values. \iFix utilizes the XStream library~\cite{noauthor_xstream_nodate} to serialize the runtime values of those variables into an XML file. 

In addition to logging those variables, \iFix also logs the runtime values of \textit{method calls and subexpressions} that use the identified variables since programmers may also wonder about their runtime values during debugging. For example, for the codebase in Figure~\ref{code-a}, \iFix logs the value of \CodeIn{n1+n2+1} in Line \#181 and the value of \CodeIn{FastMath.sqrt(VarU)} in Line \#183. 

\lstset{numbers=left, numberblanklines=false, escapeinside=||, style=code}
\begin{figure}[t]
\centering
\begin{subfigure}{0.50\textwidth}
    \begin{lstlisting}[language=Java, firstnumber=94, numbersep=-10pt, xleftmargin=-1pt, basicstyle=\ttfamily\scriptsize]
    for (int pt = 0; pt < consumed; pt++) {
        pos += Character.charCount(Character.codePointAt(input, pos));   /* Buggy */
        // pos += Character.charCount(Character.codePointAt(input, 0));  /* Patch#96 */
        // pos += Character.charCount(Character.codePointAt(input, pt)); /* Patch#97 */
    }\end{lstlisting}
    \caption{}
    \label{fig:ui-preview-2-sub1}
\end{subfigure}%
\begin{subfigure}{0.50\textwidth}
  \centering
  \includegraphics[width=\linewidth]{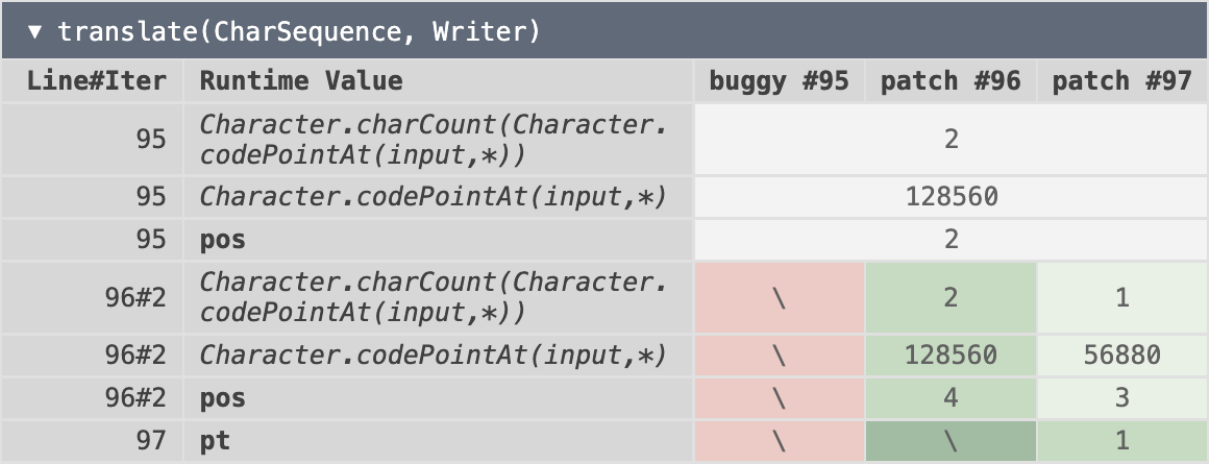}
  \caption{}
  \label{fig:ui-preview-2-sub2}
\end{subfigure}
\caption{Bug Lang 6 from Defects4J with a buggy statement in a loop (a), and the runtime comparison table \iFix generated for the Bug (b).}
\label{fig:ui-preview-2}
\end{figure}


\textbf{Runtime Value Alignment.} Finally, \iFix analyzes the log files and aligns the runtime values in a series of tables. Most runtime values can be aligned by their locations in the source code. A special case is the runtime values logged from a variable in a loop or a method called multiple times during the test execution. 
In such cases, the log files will contain multiple runtime values for the variable in the same location. For example, in Figure~\ref{fig:ui-preview-2-sub1}, the variable, \CodeIn{pos}, will be logged multiple times in the loop. Furthermore, the number of logged values of the same variable may vary across different patches, since the patches may alter the loop iterations differently. To handle such cases, \iFix assigns an index to each runtime value of the same variable based on their appearance in the log files. Then, it aligns these values based on the indices. For runtime values that are logged many times across iterations, \iFix identifies the iteration where the behavioral divergence occurs for the first time (i.e., the runtime values of the same variable differ from the buggy code and the patch) and renders the runtime values at that iteration to programmers to avoid overwhelming them.

\vspace{-0.5em}

\subsection{Implementation Details}
\label{sec:implementation-details}
\noindent We have implemented \iFix as a VS Code extension. We integrated CURE~\cite{jiang_cure_2021} as the default APR tool in \iFix, since it is open-sourced and reproducible. It was also the SOTA approach at the time of the tool development of \iFix. CURE can be replaced with any APR tool since \iFix is designed to be agnostic to any APR tools. We have also experimented with other APR techniques, including RewardRepair \cite{rewardrepair} and KNOD \cite{knod}, in the evaluation (Section \ref{sec:generalizability}). 

The front-end of \iFix is written in TypeScript and comprises \iFixFrontEndLOC lines of code. We implemented our {\em runtime behavior comparison} method in Java with \iFixInstruLOC lines of code and the {\em patch clustering and sampling} algorithm in Python with \iFixBackEndLOC lines of code. \credit{The functionality of \iFix is currently limited to fixing bugs with single-hunk patches.} Section \ref{subsec:limitations} provides a more detailed discussion. \credit{Our source code has been made publicly available on GitHub \cite{ifix_github} and Zenodo \cite{ifix_zenodo}.}



\section{User Study}
\label{sec:UserStudy}
\noindent To evaluate the overall usefulness and usability of \iFix, we conducted a within-subjects user study\footnote{\label{note_user_study}This study has been approved by the IRB office at Purdue University.} with \NumUserStudyParticipants participants. We investigate the following research questions:

\begin{itemize}[leftmargin=10mm]
    \setlength\itemsep{0.5mm}
    \item[\textbf{RQ1}] How effectively can \iFix help improve programmers' performance in validating auto-generated patches?
    \item[\textbf{RQ2}] To what extent does \iFix improve programmers' confidence on auto-generated patches? 
    \item[\textbf{RQ3}] How does \iFix help programmers with different levels of programming expertise?
\end{itemize}


\subsection{Participants}
\Comment{This is too short. You need to check my previous papers.}
\noindent We recruited \NumUserStudyParticipants programmers (4 female and 24 male) from academia and industry. 20 of the participants were recruited through mailing lists of the Department of Computer Science at Purdue university, including 14 graduate students and 6 undergraduate students. The other 8 participants were professional developers from the industry, who were recruited via word of mouth. All participants have at least 2 years of programming experience, among which 12 have 2-5 years, and 16 have over 5 years. 
In terms of Java programming, 3 of our participants have over 5 years of expertise, 10 have between 2 and 5 years, and 15 have no more than 1 year.
Furthermore, 26 participants have been using VS Code for the majority of their programming tasks for at least one year. As compensation for their participation, each participant received a \$25 Amazon gift card.

\begin{table}[t]
  \footnotesize
  \caption{Four bugs used for the user study}
  \vspace{-0.5em}
  {%
  \begin{tabular}{l|llcc}
    \toprule
     \multicolumn{1}{c|}{\multirow{2}{*}{Bug ID}} &  \multicolumn{1}{c}{\multirow{2}{*}{Project Name}} &  \multicolumn{1}{c}{\multirow{2}{*}{Root Cause}} & \multirow{1}{*}{Number of} & \multirow{1}{*}{Number of}  \\
    & & &  \multirow{1}{*}{Plausible Patches}  &  \multirow{1}{*}{Clusters}\\ 
    \midrule
    Math 30 & Apache Commons Math & Integer overflow & \multicolumn{1}{r}{33\ \ } & \multicolumn{1}{r}{5\ \ } \\
    Lang 6 & Apache Commons Lang & Wrong argument  & \multicolumn{1}{r}{7\ \ } & \multicolumn{1}{r}{2\ \ }  \\
    Chart 9 & JFreeChart & Wrong condition & \multicolumn{1}{r}{22\ \ } & \multicolumn{1}{r}{4\ \ }  \\
    Math 94 & Apache Commons Math & Integer overflow & \multicolumn{1}{r}{9\ \ } & \multicolumn{1}{r}{2\ \ }  \\
    \bottomrule
  \end{tabular}%
  }

  \label{tab:user-study-tasks}
\end{table}


\subsection{Tasks}
\label{sec:tasks}
\noindent We chose Defects4J \cite{just2014defects4j}, a widely used APR benchmark with real bugs from popular Java projects, to select bugs to repair in the user study. As described in Section \ref{sec:implementation-details}, \iFix used CURE \cite{jiang_cure_2021} as the default APR tool. Thus, we first filtered out bugs that CURE cannot generate multiple plausible patches, leaving 31 bugs. We randomly sampled 4 bugs out of the 31 bugs as the user study tasks.

Table~\ref{tab:user-study-tasks} shows the details of the sampled bugs. These bugs come from popular open-source repositories on GitHub and cover different kinds of root causes that programmers may encounter in real-world scenarios, including wrong argument, integer overflow, and wrong condition. \credit{While two bugs share the same root cause, we chose not to alter the random sampling result to ensure randomness and avoid subjective biases. Furthermore, while the root cause is the same, the two buggy programs are very different---one is about computing the greatest common divisor and the other is about computing the result of a statistical test.} Please refer to the details of each bug on our user study website.\textsuperscript{\ref{note2}}

\credit{We chose four bugs instead of more because we needed to ensure sufficient trials for each bug in each condition to measure the statistical significance of our results. Otherwise, the experiment results on a bug may largely depend on who was assigned to the bug under the condition, which makes it unreliable to compare user performance between bugs under the same condition. This is a common practice in other user studies on program repair. For example, InParFer \cite{liang_interactive_2021} used 4 bugs, and Lin et al.~\cite{lin2017feedback} used 3 bugs.}


\subsection{Comparison Baselines}
\noindent We chose two comparison baselines. The first baseline is a state-of-the-art interactive patch filtering technique---\textit{InPaFer}~\cite{liang_interactive_2021}. Among existing approaches~\cite{gao2020interactive, bohme_2020_human, liang_interactive_2021}, InPaFer is the most recent and open-source approach. It filters candidate patches by asking questions to the developers. Users can select the correct location, execution trace, and return value of the buggy method to filter out incorrect patches. Based on users' responses, InPaFer eliminates candidate patches that do not match the correct outcome. For our user study, we utilized the official implementation provided by the authors of InPaFer.

To simulate a scenario without interactive support, we design a second baseline. In this setup, participants are presented with a complete list of plausible patches generated by the underlying APR technique and are asked to identify the correct patch through manual inspection. Given that most APR approaches are developed without incorporating interactive features, we seek to assess how effectively users can navigate and select the correct patch when only the raw set of plausible patches is available. This allows us to understand the challenges users face when relying solely on traditional, non-interactive APR outputs.


\subsection{User Study Protocols}
\noindent For each participant, we selected three out of the four bugs and asked the participant to fix them in three settings: 1) Manually inspecting the list of plausible patches generated by the default APR tool; 2) Inspecting the plausible patches with \iFix; 3) Inspecting the plausible patches with InPaFer. To faithfully represent the real-world usage scenario, participants were allowed to freely edit the code, navigate the codebase, run test cases, search online, and use the default debugger integrated into the IDE in all three settings. 

\credit{To mitigate response bias \cite{FURNHAM1986385}, we followed well-established practices \cite{Bergen2020Everything} in the design and execution of our user study. We made sure none of the participants knew the experimenters or had any form of affiliation with the research group. Specifically, while the industry participants were referred by colleagues of the experimenters, we only invited those who did not know the experimenters before. During the tasks, we kept the tools anonymous and did not disclose which tool was developed by us. After the task session began, the experimenters solely observed the participants' behavior without providing any hints or guidance.}

Both task assignment and tool assignment were \textit{counterbalanced} to mitigate the learning effects. Participants were equally distributed to each combination of debugging tool assignment and task assignment. For each bug-tool pair, 7 participants worked on that bug using that tool. The ordering is randomized to mitigate the learning effect. 

At the beginning of each session, we asked the participants for their permission to screen recording. Before using \iFix and InPaFer, participants were provided with a five-minute tutorial to learn how to use them. Then, participants were given \textit{20 minutes} to use the assigned tool to fix the bug. 
A task was considered failed if participants did not manage to choose \credit{the patch that was \textit{semantically equivalent to the correct one}} after 20 minutes or if they provided an incorrect patch. At the end of each task, we asked participants to complete a post-task survey, which includes five NASA Task Load Index (TLX) questions~\cite{HART1988139} for assessing the cognitive load (Table \ref{tab:nasa-tlx}), as well as two Likert-scale questions to rate the usefulness of the tool and their confidence on the patch selected using the tool. The survey also included open-ended questions about what they liked or disliked about the assigned tool and which tool they preferred to use in practice. After all three tasks, participants were asked to complete a final survey to directly compare the three debug settings in terms of their preference and confidence about the selected patches. We have uploaded all user study data and survey forms to the GitHub repository \cite{ifix_github}.

\begin{table}[]
\footnotesize
  \caption{NASA TLX Questions for Cognitive Load Assessment}
  \vspace{-0.7em}
  {%
  \begin{tabular}{l}
    \toprule
    NASA Task Load Index Questions   \\
    \midrule
    Q1. How mentally demanding was using this tool? \\
    Q2. How hurried or rushed were you during the task?  \\
    Q3. How successful would you rate yourself in accomplishing the task?  \\
    Q4. How hard did you have to work to achieve your level of performance?  \\ 
    Q5. How insecure, discouraged, irritated, stressed, and annoyed were you? \\
    \bottomrule
  \end{tabular}%
  }
  \label{tab:nasa-tlx}
\end{table}




\subsection{Results}
\subsubsection{Task Success Rate and Completion Time}\hfill
\vspace{1mm}

\label{subsec:performance}

\begin{wrapfigure}{r}{0.32\textwidth}
    \vspace{-2em}
    \centering
    \includegraphics[width=\linewidth]{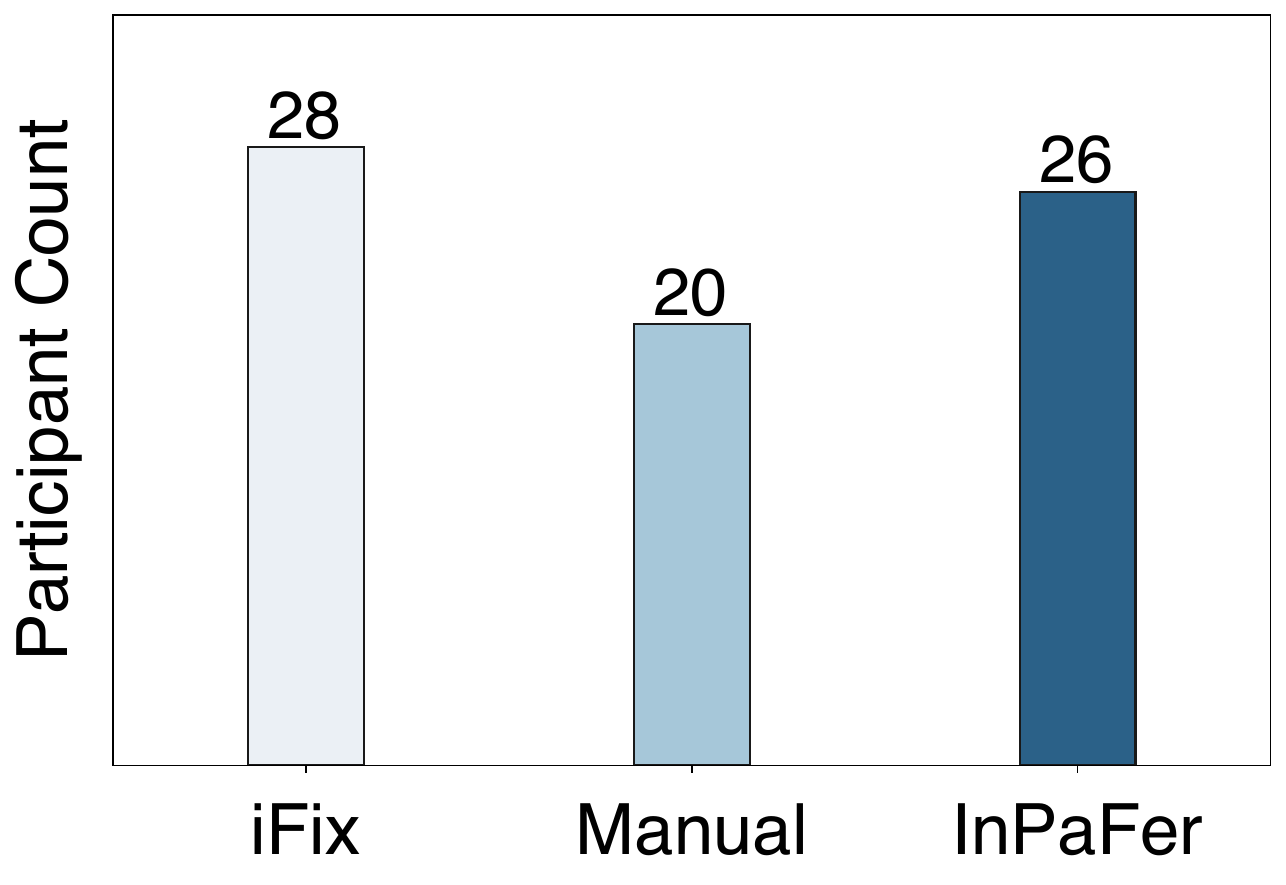} 
    \caption{Task Success Rate}
    \vspace{-1em}
    \label{fig:user_completion}
\end{wrapfigure}

\noindent As shown in Figure \ref{fig:user_completion}, when using \iFix, all \NumUserStudyParticipants participants successfully selected the correct patch from the plausible patches and fixed the bug within the given time. By contrast, 20 participants selected the correct patches in the manual condition and 26 participants selected the correct patches when using InPaFer. Compared to these two baselines, \iFix helped improve the success rate by \UserSuccessRateBoostAgainstManual and \UserSuccessRateBoostAgainstBaseline, respectively. 

Table~\ref{tab:user-study-settings} shows the average task completion time. \iFix consistently reduced the task completion time for all four bugs com-{\parfillskip0pt\par} 

\begin{wraptable}{r}{0.32\textwidth}
    \centering
    \caption{Average Task Completion Time (in Minutes)}
    \resizebox{\linewidth}{!}{
    \small 
    \begin{tabular}{c|ccc}
    \toprule
    {Bug ID} & {\xspace\xspace\textit{\iFix}\xspace\xspace} & {\textit{Manual}} & {\textit{InPaFer}}  \\
    \midrule
    Math 30 & 7.38 & 10.77 & 11.04 \\
    Lang 6 & 8.24 & 14.25 & 13.50 \\
    Chart 9 & 8.23 & 12.49 & 11.44 \\
    Math 94 & 6.21 & 9.50 & 8.55 \\
    \hline
    Average & 7.51 & 11.75 & 11.13 \\
    \bottomrule
    \end{tabular}}
\label{tab:user-study-settings}
\end{wraptable}

\noindent pared with manual inspection and using InPaFer by \UserDebugTimeReducedAgainstManual and \UserDebugTimeReducedAgainstBaseline, respectively. 
\credit{Our multivariate analysis shows that the difference in task completion time across the three conditions is significant (one-way ANOVA:  $F = 6.2138$, $p = 0.0031$). Our post hoc analysis shows that participants using \iFix completed tasks significantly faster than those using InPaFer (Tukey’s HSD test: $p = 0.0181$) and manual inspection ($p = 0.0045$), while no significant difference was observed between InPaFer and manual inspection ($p = 0.8817$).}
According to a recent survey \cite{stripe_study}, developers spend 42\% of their working time on debugging and modifying buggy code, which brings about \$85 billion global GDP loss. Thus, the one-third reduction in task completion time achieved by \iFix has a great potential to improve developers' productivity in debugging and repairing.

To understand \textit{why} participants using \iFix performed better, we analyzed the post-task survey responses and the screen recordings. We identified two main reasons. First, we found that the runtime comparison table significantly sped up the efficiency of bug-fixing by eliminating the effort of manually debugging. \credit{Based on the recordings, all 28 participants using \iFix made heavy use of the runtime comparison tables. On average, the participants interacted with this feature 2.93 times. Specifically, they clicked on variables, expressions, and subexpressions 0.46 times to navigate to their definitions; they expanded or collapsed tables 0.50 times to reduce information overload. Besides, they interacted with the interactive patch exploration feature 0.75 times by clicking on the ``Explore Similar Patches'' button to investigate more patch candidates. Only three users clicked on the ``Exclude Similar Patches'' button to reject a cluster of patches. We suspect this is because participants were concerned that they would miss the correct patch by simply rejecting a cluster of patch candidates based on a single patch.} 

\begin{figure*}[b]
    \centering
    \includegraphics[width=\linewidth]{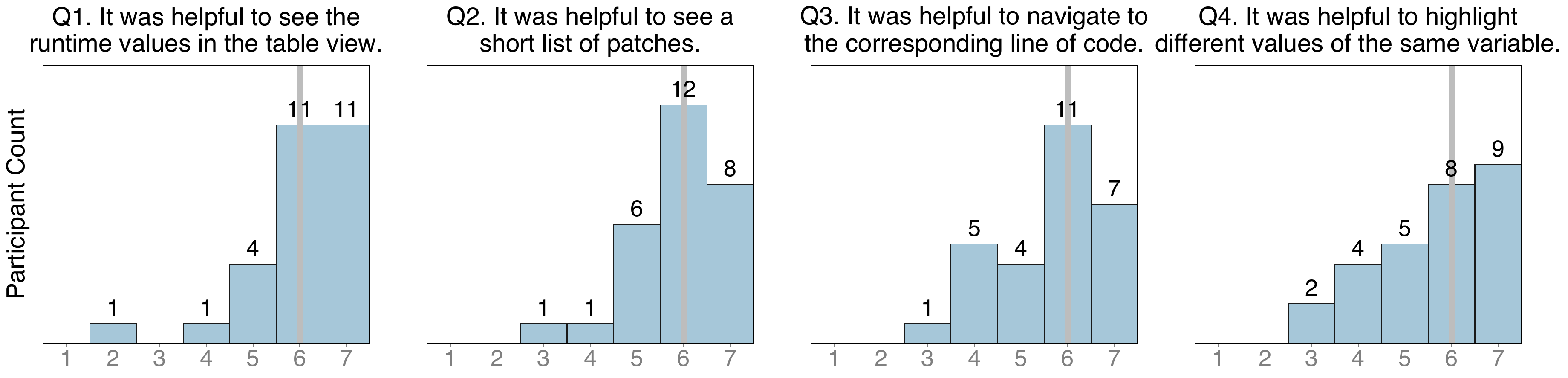}
    \vspace{-1em}
    \caption{User ratings on the usefulness of individual features (vertical bar indicating the median)}
    \label{fig:feature_rating}
\end{figure*}

Notably, most participants (27/28) sorely relied on {\iFix} to understand and validate patches without even using the IDE's debugger. This implies that the runtime comparison tables provide sufficient information for understanding the runtime behavior of patch candidates. During inspection, they navigated between different methods and files by clicking on the variable names displayed in the table. By contrast, when using InPaFer, 5 participants still used the debugger to run the program and determine the correct answers to InPaFer's questions, while others tended to examine the source code line by line without running the program. P15 said, ``{\em I think it (InPaFer) needs the user to have a correct calculation when debugging.}'' Under the manual inspection setting, participants relied entirely on their own strategies to verify the candidate patches. 9 participants utilized the debugger, setting breakpoints, checking runtime values step-by-step, and comparing the behavior of different candidate patches. The other 19 participants took the strategy of eye-balling the code line by line to trace key variables. While 12 of them utilized the integrated widgets in the VSCode IDE for navigation, the remaining 7 relied on searching for specific methods and variable names, which slowed down the debugging process. In our post-task survey, 22 of our participants agreed or strongly agreed that being able to see the runtime values of different patches side by side in the table view helped them recognize the correct patch  (Figure \ref{fig:feature_rating}, Q1). P26 said, ``{\em The view in VSCode where variable value and different values for intermediate states were crucial in understanding what each patch does and whether it would behave as expected.}''

Second, \iFix significantly reduces the number of candidate patches they need to inspect per bug. When using \iFix, participants heavily utilized the ``Explore Similar Patches'' button (Figure~\ref{teaser-1a}) to dive into clusters of patches similar to their preferred candidates. Specifically, 6 participants clicked this button twice to explore two clusters of patch candidates, 17 participants clicked once, while the other 5 managed to identify the correct patch from the initial set of representative candidates selected by \iFix without further exploration. In contrast, when using InPaFer, we discovered that even after participants had answered all questions prompted by the tool, there were still an average of 8.5 patches per bug that could not be further filtered. Since InPaFer only supports displaying 3 candidate patches at once, they needed to click on a button to exclude the undesired candidates one by one. P25 said, ``{\em It (InPaFer) doesn't show all candidates at once.}'' Under the manual inspection setting, 13 participants followed the APR tool's ranking to inspect the ranked list of candidate patches, comparing different candidates as they went through the list. The other 15 participants began by selecting a patch based on their intuition and then compared it with other less-preferred candidates. Such a process became increasingly overwhelming as more patches were added to the comparison. P4 said, ``{\em Reading lots of patches may distract me. I just need more information.}'' P24 said, ``{\em Some of the candidates are quite misleading, making it hard for me to make a choice.}'' In our post-task survey, 20 out of 28 participants agreed or strongly agreed that being able to see a short list of representative patches, explore similar patches, and exclude unpromising patches helped them navigate the generated patches and identify the correct one (Figure \ref{fig:feature_rating}, Q2). P9 said, ``{\em I like the group feature (of \iFix) so I only need to review representative patches.}''

\credit{One may wonder how well \iFix works with cases where textually similar patches behave very differently. When an incorrect patch and a correct patch are textually similar, they may be clustered together and the incorrect one may be selected as a representative patch. This happens in one-fourth of the cases. However, we noticed that participants barely directly rejected a cluster of patches just based on the representative patch. They often first zoomed into the cluster and glanced over other similar patches first. Furthermore, participants did not only make decisions based on the clustering result or the text similarity between patches. All of them checked the runtime behavior table which highlights the runtime behavior differences between patches via color-coding. As a result, they quickly found the correct patch buried in a cluster. This explains why all 28 participants quickly identified the correct patch, even when it was not the cluster’s representative in \iFix.}

\credit{We noticed some differences between the results of our user study and those reported in the InPaFer paper. For the bug {\em Chart 9}, while the InPaFer paper reported a significantly shorter task completion time compared to ManualFix \cite{liang_interactive_2021}, Table \ref{tab:user-study-settings} shows only a minor difference. This difference stems from the use of different sets of candidate patches in the experiments. 
In the user study in the InPaFer paper, they used a combination of plausible patches generated by traditional APR tools \cite{liang_interactive_2021}. These patches differ significantly from each other and can be effectively filtered by answering InPaFer's questions. 
In our study, the plausible patches are generated by CURE \cite{jiang_cure_2021}, an advanced learning-based tool. Many of the patches only have subtle differences from each other, and they cannot be further filtered by answering InPaFer's questions. 
Given that InPaFer's questions are pre-defined and only examine certain behaviors, participants had to manually review the remaining patches. As noted earlier, even after answering all the tool’s questions, an average of 8.5 patches per bug remained unfiltered and required manual inspection and comparison. As a result,  their performance showed little improvement over manual patch selection.}

\begin{finding}{Impact of \iFix on User Performance}
Compared to manually inspecting patches and using another patch filtering technique, \iFix significantly improves the task success rate by 40\% and 8\%, respectively, while also reducing task completion time by 36\% and 33\%.
\end{finding}

\subsubsection{User Confidence and Cognitive Overhead}
\begin{figure}[t]
    \centering
    \footnotesize
    \includegraphics[width=\columnwidth]{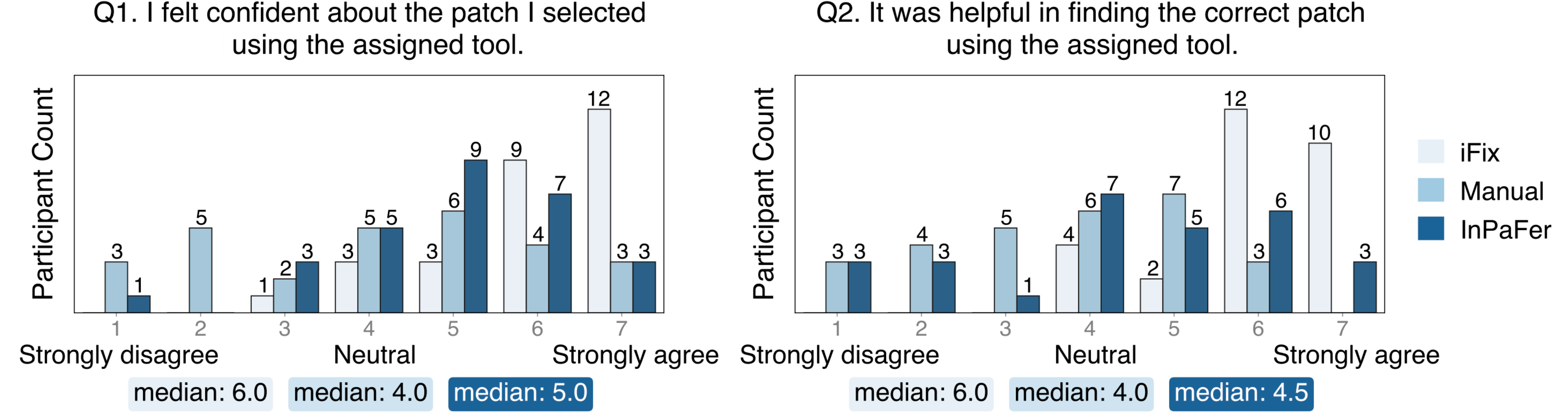}
    {(a) \hspace{175pt} (b) \hspace{35pt}}
    \vspace{-1em}
            \caption{The distribution of participants’ confidence in the patches selected using the assigned tools (a) and preference on the assigned tools (b).}
    \label{fig:user_feedback}
\end{figure}

\hfill
\vspace{1mm}
\\
\label{subsec:confidence}
\noindent\iFix significantly enhances developers' confidence in auto-generated patches. In the post-task survey, 21 participants agreed or strongly agreed that they felt confident about the patches when using \iFix, as shown in Figure \ref{fig:user_feedback}a. In contrast, only 7 and 10 participants felt so when inspecting patches manually and when using InPaFer. The mean difference of the users' confidence on the patches selected using \iFix is statically significant compared to manual inspection (df = 27, p-value = 0.00021; \UserConfidenceBoostAgainstManual increase on median) and InPaFer (df = 27, p-value = 0.00404; \UserConfidenceBoostAgainstBaseline increase on median). 

This improvement was largely attributed to a better understanding of the runtime behavior of candidate patches. In the post-task survey, we asked the participants why they felt more confident about the tool they chose in an open-ended question. 15 participants mentioned that \iFix's runtime comparison table helped them identify the root cause of the bug and how each patch attempted to fix it. P8 said, ``{\em (I feel more confident) because the information shown in \iFix helps me understand what is going on and I know the mechanism.}'' P16 said, ``{\em Because I basically know where the problem is offered with different variables information.}'' Interestingly, even after noticing the correct patch, 21 participants manually inspecting patches and 16 using InPaFer spent an average of 3.4 minutes comparing it with other candidates. P9 said, ``{\em At the end of this experiment, I am questioning whether the generated patches are correct. In addition, I cannot responsibly accept a patch that I did not understand, so I have to rethink what the patch actually does anyway.}'' Besides the contribution of the runtime comparison table, 2 participants also highlighted that color-coding different runtime values in the runtime comparison table also helped. P21 said, ``{\em (I feel more confident) Because I can easily select the right patch according to the color and output of the runtime stack, just pick the patches with the desired output and think a little bit then the job is done.}''

\begin{figure}[h]
    \centering
    \includegraphics[width=0.57\columnwidth]{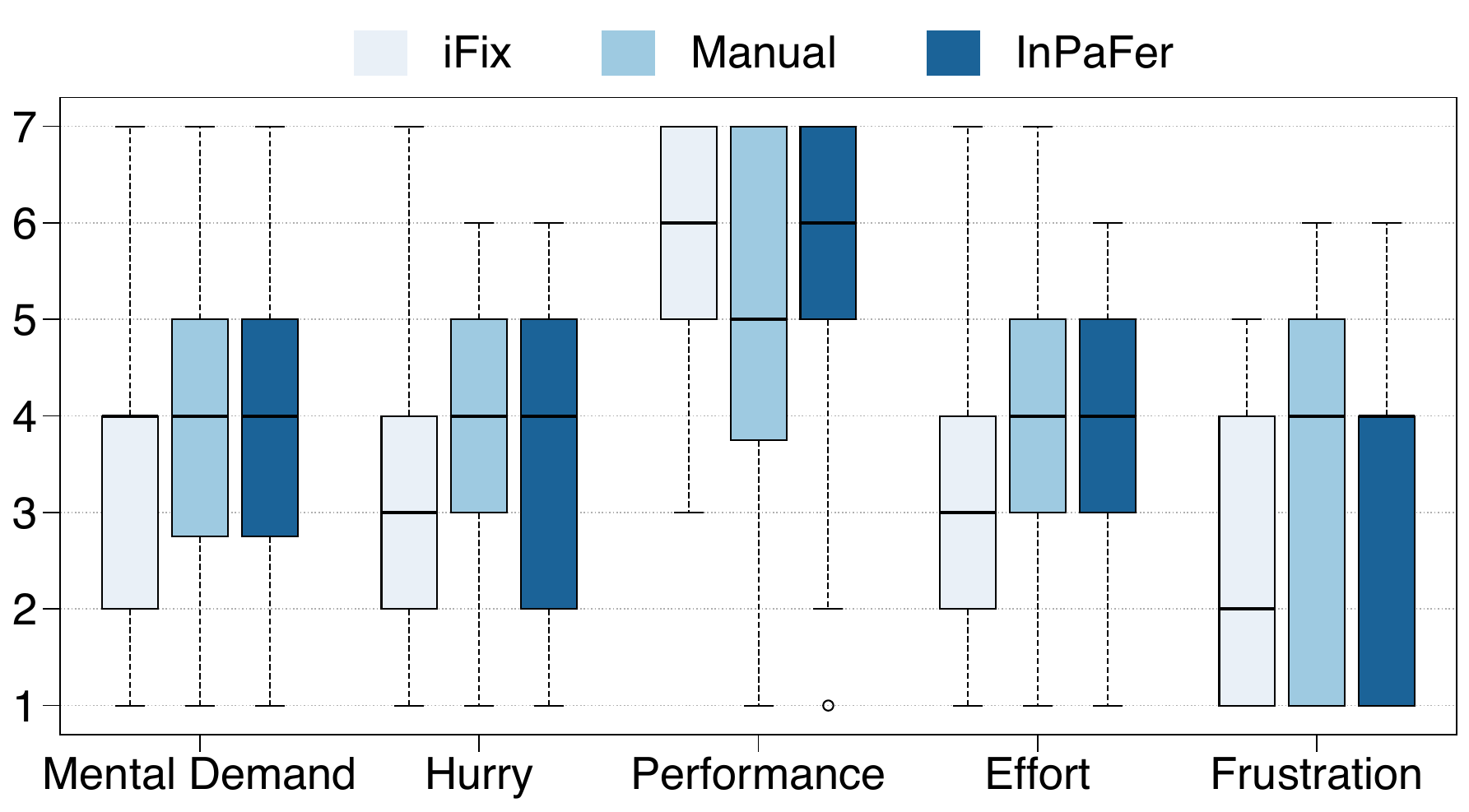}

    \caption{Cognitive load measured by NASA TLX~\cite{HART1988139}}
    \Description{Cognitive load measured by NASA TLX. Participants felt less mental demand, hurry, effort, and frustration when using \iFix.}
    \label{fig:nasa_tlx}
\end{figure}

Since \iFix has more sophisticated features than InPaFer and a traditional IDE, one may wonder if \iFix leads to more mental demand. After each task, participants self-reported the cognitive load of using the assigned tool in a NASA TLX questionnaire~\cite{HART1988139}. As shown in Figure ~\ref{fig:nasa_tlx}, participants felt less mental demand, hurry, and frustration when using \iFix. As mentioned in the previous paragraphs, \iFix's runtime comparison table eliminated the effort of manual debugging. In addition, \iFix presents the selected candidate patches directly below the modified code, allowing users to easily compare them without navigating between different editors. P26 said, ``{\em Seeing a small set of patches inline was less overwhelming to identify the right patch.}'' Last but not least, \iFix reduces the number of candidate patches for inspection, making users feel less overwhelmed. P4 said, ``{\em It (\iFix) reduces the number of patches so I can more focus on the given patches and think about them.}''

\begin{finding}{Impact of \iFix on User Confidence}
Compared to manually inspecting patches and using another patch filtering technique, using \iFix can significantly enhance users' confidence in the patch they select by 50\% and 20\% on average.
\end{finding}

\subsubsection{User Preference and Feedback}
\hfill
\vspace{1mm}
\\
\noindent In our post-task survey, 22 of \NumUserStudyParticipants participants agreed or strongly agreed that \iFix was helpful in finding the correct patch, as shown in Figure \ref{fig:user_feedback}b. By contrast, only 9 of them agreed or strongly agreed that InPaFer was helpful, and 3 agreed or strongly agreed that manual inspection was helpful. The mean difference is statically significant compared to manual inspection 
(paired t-test: df = 27, p-value < 0.00001; \UserPreferenceBoostAgainstManual increase on the median) and also to InPaFer (df = 27, p-value = 0.00030; \UserPreferenceBoostAgainstBaseline increase on the median). We coded participants’ responses to the open-ended question about their experience with \iFix. We identified two themes. First, 19 of our participants highlighted that \iFix helped them validate and compare different patch candidates easily. P20 said, ``{\em It (\iFix) can display the value of some key variables(relate to branch conditions) in each line so I can analyze the branch conditions quickly.}'' P23 said, ``{\em Otherwise I need to go through the codes and print intermediate values.}'' Second, 6 participants pointed out that with the aid of \iFix, they do not need to examine too many patches. P24 said, ``{\em The similar candidates were grouped [by \iFix], making it easier for me to find the initial thought to solve the issue.}''

We also asked participants what additional features may help them better solve the tasks. 3 participants mentioned that \iFix could explain the runtime behavior in a better way, like a workflow chart. 4 participants suggested that \iFix could summarize the runtime behavior of different patches and their differences in natural language. Finally, 2 participants expected more interaction with \iFix, e.g. one may want to manually add/edit patches and inspect the runtime values.

\begin{figure}[t]%
    \includegraphics[width=0.52\linewidth]{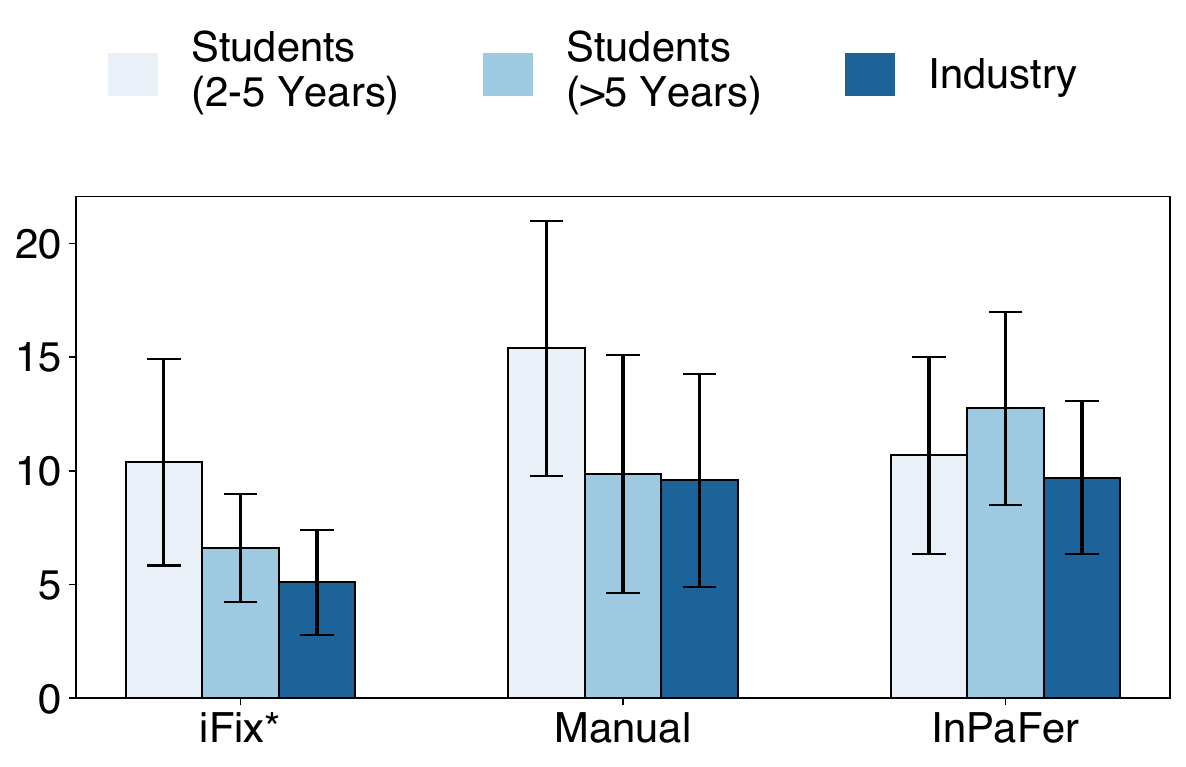}
    \caption{Average task completion time (in minutes) of different groups of participants, a star(*) indicates the mean difference is statistically significant.}%
    \label{fig:u5-a5-ind}
\end{figure}

\subsubsection{Impact of Programming Expertise}
\hfill
\vspace{1mm}
\\
\label{sec:impact_of_expertise}
\noindent The performance of \iFix users was affected by their programming expertise. For example, experienced programmers tend to resolve bugs more quickly than novices. Figure~\ref{fig:u5-a5-ind} shows the impact of programming experience on participants with different programming expertise, \credit{including students with 2–5 years of experience, students with more than 5 years of experience, and industry professionals. Our multivariate analysis confirms that programming expertise significantly influenced participants’ performance across the three debugging methods (MANOVA, Wilks’ lambda = $0.4984$, $p = 0.0105$). Further analysis revealed that the difference in programming expertise was particularly significant when using \iFix, where participants’ debugging time varied significantly across occupation groups (ANOVA, $p = 0.0097$). In contrast, while manual debugging showed a marginally significant difference ($p = 0.0503$), no significant difference was observed under the InPaFer setting ($p = 0.3133$). These results highlight that \iFix effectively leverages participants’ programming expertise, enabling more experienced users to debug more efficiently.}

Compared to students with 2-5 years of programming experience, those with more than 5 years of experience performed significantly better under the condition of using \iFix (36.4\%, Welch's t-test: p-value $= 0.041$) and manual inspection (35.9\%, Welch's t-test: p-value $= 0.044$). Both groups of students performed better using \iFix compared to manual inspection and using InPaFer. 
Compared to using \iFix, student participants with greater than 5 years of programming experience spent 93.0\% more time using InPaFer and 49.4\% more time with manual debugging, while the mean difference between using \iFix and using InPaFer is statistically significant (paired t-test: p-value $= 0.001$). 
Compared to using \iFix, Those with 2-5 years of programming experience spent 48.4\% more time manually debugging and similar time with InPaFer.

Surprisingly, when using InPaFer, experienced students took 19.3\% more time to complete the assigned program repair tasks. Yet the mean difference is not significant (Welch's t-tests: p-value $= 0.16$). 
We revisited our video recording to figure out the reason. We found that experienced students often strived to explore more patch candidates to find the best patch, while InPaFer is not convenient to explore a large number of patches and thus costs them more time. 

\begin{finding}{Impact of programming expertise}
As programming expertise increases, the performance of using \iFix improves significantly. Participants with over 5 years of programming experience showed a 36.4\% improvement compared to those with 2-5 years. Industry professionals completed bug repairs 40.0\% faster than student developers.
\end{finding}

\section{Quantitative Experiments}
\label{sec:QuantitativeExperiment}
\noindent \credit{The user study has evaluated the usefulness of \iFix as a holistic approach.} Since \iFix employs the patch clustering and sampling algorithm to select a small set of patches as a starting point and re-rank the representative patch candidates, one may wonder how good these initial patches are. Thus, we conducted additional experiments to quantify the effectiveness of our patch clustering and sampling algorithm. We investigate the following research questions:

\begin{itemize}[leftmargin=10mm]
    \setlength\itemsep{0.5mm}
    \item[\textbf{RQ4}] Compared with other patching ranking methods, how effective is the patch clustering and sampling algorithm in terms of the final rank of the correct patch? 
    \item[\textbf{RQ5}] How sensitive is the patch clustering and sampling algorithm to the underlying APR techniques? 
\end{itemize}

\subsection{Comparison Baselines and Experimental Setup}
\noindent To answer RQ4, we selected three different kinds of patch ranking methods as our baselines: (1) the original patch ranking given by the APR tool, (2) a state-of-the-art static patch ranking method~\cite{le2017s3}, and (3) a learning-based re-ranking method adopted by a state-of-the-art APR tool, AlphaRepair~\cite{xia_less_2022}.  We elaborate on each comparison baseline below. 

\begin{itemize}[leftmargin=*]
    \item \textbf{Original Patch Ranking}: We use the default ranking obtained from CURE~\cite{jiang_cure_2021}. Specifically, CURE ranks patches based on the probability score of each patch.
    \item \textbf{Static Patch Ranking}: According to a recent comparative study on static patch ranking methods~\cite{wang2020automated}, S3~\cite{le2017s3}  achieves the best performance in ranking correct patches compared with other static methods proposed in~\cite{wen2018context, xin2017leveraging, tan2016anti}. Therefore, we use S3 as the static patch ranking baseline. Specifically, S3 uses three static code features: {\em Abstract Syntax Tree (AST)-level edit distance}, {\em cosine similarity computed from ASTs}, and {\em the locality of variables and constants}.
    \item \textbf{Learning-based Patch Ranking}: Recent patch ranking techniques \cite{xia_less_2022, shibboleth} leverage advanced  learning models to rank patches. We include the latest one, AlphaRepair \cite{xia_less_2022} as our learning-based patch ranking baseline. AlphaRepair masks tokens in a patch, adopts a pre-trained large language model, CodeBERT~\cite{feng2020codebert}, to predict the probability of the masked token, and then uses the average probability of all tokens as the ranking score. 
    \item \textbf{Similarity-based Ranking} (Ablation Study): We create a variant of our algorithm \iFix\textsubscript{S} by ablating the clustering step to understand its contribution to our method. For each plausible patch, this variant only calculates the Levenshtein distance to the target buggy code and ranks the patches based on it. 
     
\end{itemize}

Following the user study, we continued to use CURE~\cite{jiang_cure_2021} as the default APR technique. we used the 31 Defect4J bugs for which CURE can generate multiple plausible patches. For each bug, we applied our patch clustering and sampling algorithm and the baseline patch ranking approaches to the patches generated by CURE and compared the final ranking of the correct patch. 

Specifically, the ranking of a patch in a hierarchical cluster is determined by its position within the nested structure of clusters.
For example, if a patch appears in the \( n \)-th level of a hierarchical cluster and is ranked as the \( X \)-th patch within that specific sub-cluster, and there are \( \{a_1, a_2, \ldots, a_{n-1}\} \) clusters at each level along the path leading to it, then the final rank of the patch is calculated as:
\begin{align*}
    a_1 + a_2 + \ldots + a_{n-1} + X
\end{align*}
This ranking strategy ensures that patches are prioritized by their positions across the entire hierarchical structure, taking into account both their level within the hierarchy and their relative positions within their respective sub-clusters.

To answer RQ5, we chose two additional APR techniques as the baselines: RewardRepair~\cite{rewardrepair} and KNOD~\cite{knod}. Both RewardRepair and KNOD are recent, open-sourced, and state-of-the-art APR techniques. 
Similar to the previous experiments, we used  Defects4J~\cite{just2014defects4j} as the benchmark. Following the previous criteria, we focused on bugs for which these APR techniques generate multiple plausible patches. We obtained 32 bugs for RewardRepair, 31 bugs for CURE, and 43 bugs for KNOD. In this experiment, we applied our patch clustering and sampling algorithm to the patches generated by each of these APR tools and compared the final ranking of the correct patches against their original rankings produced by the respective APR techniques.

\subsection{Results}
\subsubsection{Effectiveness of Patch Clustering and Sampling Algorithm}
Table~\ref{table: comparison results of patch ranking} shows the comparison results of different patch ranking methods. Overall, our patch clustering and sampling algorithm outperforms all comparison baselines, with an average ranking position of 1.7. 
Our method improves the overall ranking of correct patches by~\CUREBoost compared to the original ranking of CURE~\cite{jiang_cure_2021} and surpasses the other two methods by~\SRankingBoost and~\LRankingBoost, respectively. The results show the effectiveness of our multi-criteria method in improving the rank of correct patches.

Our ablation study shows that without hierarchical clustering, 
the performance of \iFix is reduced by 23\%. This result shows the effectiveness of our patch clustering and sampling method in improving the rank of correct patches.

\begin{table}[h]
\footnotesize
\centering
\caption{Comparison of different patch ranking methods}
\label{table: comparison results of patch ranking}
\vspace{-2ex}
{
\begin{tabular}{l|rrrrr}
 \toprule
       & \multicolumn{1}{l}{Original} & \multicolumn{1}{l}{S3~\cite{le2017s3}} & \multicolumn{1}{l}{AlphaRepair~\cite{xia_less_2022}} & \multicolumn{1}{l}{\iFix\textsubscript{S}} &  \multicolumn{1}{l}{{\bf \iFix}} \\ \midrule
MEAN & 3.5                              & 3.1                    & 25.9                             & 2.2         & {\bf 1.7}                      \\  \bottomrule
\end{tabular}}
\end{table}

\begin{finding}{Effectiveness of the patch clustering and sampling Method}
Compared to the state-of-the-art static and learning-based patch ranking methods, our patch clustering and sampling algorithm improves the overall ranking of correct patches by~\SRankingBoost and~\LRankingBoost.
\end{finding}





\subsubsection{Generalizability to Other APR Techniques}
\label{sec:generalizability}

Table~\ref{table: comparison results of APR} shows the median ranking positions of the correct patch produced by our method compared with the default rankings from three APR techniques. Our method consistently improves the default rankings by at least \MCRGenBoost compared with the three APR techniques.

\begin{table}[h]
\centering
\footnotesize
\caption{Effectiveness of our method on different APR tools}
\label{table: comparison results of APR}
\vspace{-2ex}
\begin{tabular}{l|rr|rr|rr}
 \toprule
       & {RewardRepair} & {\bf \iFix} & {CURE} & {\bf \iFix} & {KNOD} & {\bf \iFix} \\ \midrule
MEAN    & 3.4                              & \textbf{1.4}              & 3.5                      & \textbf{1.7}              & 2.8                      & \textbf{1.7}             \\  \bottomrule
\end{tabular}
\end{table}

\begin{finding}{Generalizability of the Patch Clustering and Sampling Method}
The consistent improvement of the overall ranking of correct patches shows that the {\em patch clustering and sampling} algorithm is generalizable to different APR tools.
\end{finding}

\section{Discussion}
\label{sec:Discussion}
\subsection{Design Implications} 

Our findings suggest that to help improve the productivity of developers, it is important to help them examine and compare the runtime behavior of patches. As discussed in Section \ref{subsec:performance}, without the assistance of an interactive tool, developers often spend a significant amount of time on debugging. They may set breakpoints and check runtime values step-by-step to compare the behavior of different candidate patches, or manually inspect the code line by line to trace key variables, both of which can be time-consuming. Yet, as discussed in Section \ref{subsec:confidence}, having already noticed the patch that should be correct, 75\% of the participants still spent several minutes going through the source code to confirm its correctness. Such a situation was rare when they used \iFix, when only one participant also used the debugger to complete the task. This provides strong evidence for enhancing the debugging experience in the presence of multiple plausible patches.

While \iFix provides a holistic comparison of runtime among different candidate patches, our participants provided insightful suggestions for improving \iFix. First, we can leverage large language models (LLMs) to explain the root cause of the underlying bug as well as how different patches attempted to fix it 
in natural language, which can further reduce developers' efforts. To achieve this, additional mechanisms are needed to manage potential hallucinations from LLMs. Second, we can use more advanced data visualization (e.g. flow charts) to enhance the readability of \iFix's runtime comparison tables. These suggestions highlight a potential direction for the APR community to work towards bridging the gap between developers and APR techniques.

\subsection{Threats to Validity} 
\label{subsec:threats}

Regarding \textit{internal validity}, the performance of participants in our user study may have been influenced by their programming skills, as well as the difficulty of their task assignments. 
To mitigate this threat, we selected participants with diverse levels of programming experience and chose four different bugs with a wide range of root causes from Defects4J. We also counterbalanced the task assignment to reduce the learning effect in a within-subjects study. 

While \iFix attempts to reduce software engineering costs for fixing real-world software bugs, \credit{the majority of participants in our user study were university students. 
While previous research has highlighted that CS students can serve as reasonable stand-ins for developers in the context of user studies related to software engineering~\cite{tahaei2022recruiting, Ko2015pratical}, a larger-scale study involving industry professionals would be valuable.}

\credit{The differences in interface design between \iFix and InPaFer may influence user productivity, potentially introducing a confounding factor in our user study results. While this could have some effect, we expect it to be small. Like \iFix, the user interface of InPaFer was also carefully designed and perceived as easy to use in their user study \cite{liang_interactive_2021}. Thus, we believe the users' productivity was not significantly affected by the learnability or aesthetics of the tools.}

Regarding \textit{external validity}, \iFix is primarily designed to fix Java bugs, while some APR techniques are capable of fixing bugs in multiple programming languages~\cite{xia_less_2022}, such as Python and JavaScript. This issue can be further addressed by developing \iFix for different target programming languages. 

\credit{\iFix's user study uses patches generated by CURE, an APR tool released in 2021. One may question its generalizability to more advanced APR tools, including LLM-based approaches that have shown significant improvements in patch generation. Recent studies~\cite{yang2024revisitingunnaturalnessautomatedprogram, yang2023large} have revealed that the issue of plausible but incorrect patches (i.e. overfitting)  remains a challenge even for advanced APR techniques. On Defects4J, 45\% of CURE’s patches are overfitting~\cite{jiang_cure_2021}, while AlphaRepair~\cite{xia_less_2022} and ChatRepair~\cite{xia2024automated} produce 32\% and 30\%, respectively. Nevertheless, it is worthwhile to study the effectiveness of \iFix with patches generated by more advanced APR tools.}

Regarding \textit{construct validity}, the performance of \iFix may be affected by the code similarity metric chosen for patch clustering and sampling. We used Levenshtein distance in our implementation as it effectively represents code syntax differences. Further experiments should explore the impact of using different code similarity metrics on \iFix.

\subsection{Limitations and Future Work}
\label{subsec:limitations}

\credit{As mentioned in Section \ref{sec:Approach}, \iFix logs many runtime values of variables, sub-expressions, and method calls. It is challenging to define an effective similarity metric among the sets of runtime values produced by different patches. This is because while there are many runtime values, the key semantic difference may be only reflected in a few of them. If we treat all runtime values equally, we may miss subtle differences that reflect the key semantic difference between correct and incorrect patches. Prior work \cite{xiong2018identifying, varfix} has incorporated dynamic program features into patch evaluation. Integrating this information and a broader set of metrics into interactive patch clustering is an interesting direction to explore in the future.}

\credit{Although we do not have assumptions on types of bugs, number of clusters, differences between patches, etc., \iFix is currently limited to fixing bugs with single-hunk patches.} While it is a common practice to select single-hunk bugs for research purposes~\cite{saha_elixir_2017, le_history_2016, jiang_cure_2021, xia_less_2022}, the interface of \iFix could be enhanced to help developers view and compare multiple code changes simultaneously. One solution would be to leverage the split view design to display multiple code locations in the code editor and leverage color-encoding to show the visual correspondence between different hunks of a patch in different code locations. 

\credit{If the underlying APR tool generates much longer patches, the differences can hardly be displayed in the proposed user interface of \iFix. In such cases, different patches could be displayed in separate panels that developers can switch between by clicking. In this way, developers would not be overwhelmed with multiple extensive patches cluttering the code editor.}

\section{Related Work}
\label{sec:RelatedWork}
\noindent \textbf{Interactive Support for APR}
Recently, several interactive approaches have been proposed to enhance the usability of APR techniques. 
Böhme et al.~proposed LEARN2FIX~\cite{bohme_2020_human}, a human-in-the-loop repair technique that asks programmers about the expected program output to build a test oracle before patch generation. 
Gao et al.~\cite{gao2020interactive} proposed an interactive approach that helps developers filter out undesired patches by asking questions about the program's semantics, e.g. what is the expected value of a certain variable.
Liang et al.~proposed InPaFer~\cite{liang_interactive_2021}, which helps developers filter patches by asking questions about the expected execution trace and return value of the buggy method.
As mentioned in Section ~\ref{sec:intro}, these approaches rely on pre-defined questions and are limited in scope and unable to comprehensively capture runtime behavior differences, as candidate patches often vary at specific points that such questions may fail to address. Sometimes, developers still need to run or even debug the program to correctly answer their questions. 

An alternative approach is to cluster semantically similar patches. Cashin et al. ~\cite{cashin2019understanding} used invariant sets for clustering, while Martinez et al.~proposed xTestCluster ~\cite{martinez2024test} to cluster patch candidates based on their outcome on a test oracle. However, these methods do not provide any support for navigating through the clusters and comparing patches from different clusters. Compared to these clustering methods, \iFix allows users to interactively zoom into a cluster of candidates users find promising and filter out clusters of undesired candidates. Moreover, \iFix provides a holistic view of the program's runtime behavior across multiple patch candidates. \iFix is able to provide developers with interactive support to explore patch candidates effectively, while also providing a fine-grained comparison between different patch candidates.

\vspace{1mm}

\noindent \textbf{Patch Ranking and Classification} 
Approaches for evaluating the correctness of APR-generated techniques include patch ranking and classification. 
Mechtaev et al.~\cite{directfix} and Le et al.~\cite{le2017s3} employed static analysis to rank patches based on their similarities to the original code. 
As for learning-based approaches, 
Ye et al.~\cite{ye_2021_automated} trained a model to classify correct and incorrect patches by analyzing various code features.
Long et al.~\cite{long2016automatic} and Saha et al. ~\cite{saha_elixir_2017} trained statistic models to predict patch ranking, while Ghanbari et al.~\cite{shibboleth} employed a hybrid of static ranking algorithm and classification model to prioritize correct patches. 
Xia et al.~\cite{xia_less_2022} employed a pre-trained large language model to rank patches by calculating the scores based on the joint probability of generated tokens. 

Despite these techniques, developers may still face challenges reviewing an extensive list of patches, especially when the correct patch is not highly ranked. \iFix employs a patch clustering and sampling algorithm, which improves the final ranking of correct patches compared to the state-of-the-art static and learning-based patch ranking techniques.

\vspace{1mm}

\noindent \textbf{Empirical Studies on APR}
Over the past decade, various APR techniques have been proposed. We refer the readers to recent surveys~\cite{luca2019survey, zhang2023survey} of APR for details.
There has been a growing interest in studying the usability challenges in APR. Smith et al.~\cite{smith_is_2015} and Le et al.~\cite{ le_overfitting_2018} investigated the overfitting issue in APR, while Long et al.~\cite{long_analysis_2016}, Ye et al. ~\cite{ye_2021_automated} and Lin et al. ~\cite{lin_2022_context} proposed approaches to address this issue and improve the quality of auto-generated patches. Noller et al.~\cite{noller_trust_2022}, Winter et al.~\cite{winter_lets_2023, winter_how_2022} and Zhang et al.~\cite{zhang_program_2022} conducted empirical studies and surveys to understand the developer needs and concerns about adopting APR techniques in real-world scenarios. These studies provided substantial insights for research on APR techniques. 


Meanwhile, APR techniques have been increasingly studied and adopted in the industry, 
such as Meta's SapFix \cite{harman_finding_2018}. In addition to a traditional G\&V pipeline, SapFix involves human approval to pick the best fix. Marginean et al. ~\cite{marginean_sapfix_2019} further found that developers still play an indispensable role by reviewing APR-generated bug fixings before landing them in the codebase.

\section{Conclusion}
\label{sec:Conclusion}
\noindent This paper presents \iFix, an interactive approach that allows programmers to quickly explore a large number of patch candidates and examine their behavior differences without manual debugging. \iFix consists of a {runtime behavior comparison} method to capture the runtime behavioral differences among buggy and patched code, to help programmers better understand the bug and candidate patches. \iFix also uses a patch clustering and sampling algorithm to select a small set of diverse and representative patches to reduce the number of candidate patches for programmers to inspect. 
We conducted a within-subjects user study with \NumUserStudyParticipants participants to demonstrate that \iFix effectively helps improve the performance and confidence of bug fixing. Our additional experiments demonstrated the effectiveness and generalizability of our patch clustering and sampling algorithm to different APR techniques. 


\section{Data Availability Statement}

The artifact of this paper have been made available on GitHub \cite{ifix_github} and Zenodo \cite{ifix_zenodo}. This artifact contains the source code for our VSCode plugin (detailed in Section \ref{sec:implementation-details}), the data and results from our user study (Section \ref{sec:UserStudy}), and the code and results of the quantitative experiments (Section \ref{sec:QuantitativeExperiment}).
We've also posted our user study data on a Google Site for better visualization.\textsuperscript{\ref{note2}}

\begin{acks}
We would like to thank all anonymous participants in the user study and anonymous reviewers for their valuable feedback. This work was supported in part by Amazon Research Award and the National Science Foundation (NSF Grant ITE-2333736 and CCF-2340408).
\end{acks}

\bibliographystyle{ACM-Reference-Format}
\bibliography{Reference}

\end{document}